\documentclass[aps,prb,preprint,groupedaddress,showpacs]{revtex4-1}
\usepackage{color}
\usepackage{graphicx}
\usepackage{times}
\usepackage{bm,amssymb}
\usepackage{multirow}
\usepackage{amsmath}
\usepackage{mathtools}
\usepackage{natbib}

\begin{document}

%\preprint{}

\title{Nearly Perfect Fluidity in a High Temperature Superconductor}


\author{J.D. Rameau}
\affiliation{Condensed Matter Physics and Materials Science Department, Brookhaven National Lab, Upton, NY, 11973, USA}
\email[Direct correspondence to:]{jrameau@bnl.gov}
\author{T.J. Reber, H.-B. Yang, S. Akhanjee, G.D. Gu and P.D. Johnson}
\affiliation{Condensed Matter Physics and Materials Science Department, Brookhaven National Lab, Upton, NY, 11973, USA}
\author{S. Campbell}
\affiliation{Department of Physics and Astronomy, Iowa State University, Ames, IA, 50011, USA}

%\thanks{}


\date{\today}

\begin{abstract}
Perfect fluids are characterized as having the smallest ratio of shear viscosity to entropy density, $\eta/s$, consistent with quantum uncertainty and causality. So far, nearly perfect fluids have only been observed in the Quark-Gluon Plasma (QGP) and in unitary atomic Fermi gases (UFG), exotic systems that are amongst the hottest and coldest objects in the known universe, respectively. We use Angle Resolve Photoemission Spectroscopy (ARPES) to measure the temperature dependence of an electronic analogue of $\eta/s$ in an optimally doped cuprate high temperature superconductor, finding it too is a nearly perfect fluid around, and above, its superconducting transition temperature $T_{c}$.
\end{abstract}

%\keywords{}

\maketitle
\section{Introduction}
Quantum fluids are well understood in two opposing limits: the ``collisionless" regime and the ``hydrodynamic", collision-dominated regime. They are characterized by the time between particles' collisions with each other being much longer or shorter, respectively, than the characteristic time for collisions with their surroundings\cite{kivelsonviscosity}. Generally, the cuprates have been treated pertubatively, adding the effects of interactions to a coherent collisionless system in which electronic excitations are treated as free carriers with basic properties renormalized by interactions, the quasiparticle approach. But this method has failed to completely account for the interesting behavior of electrical transport in the cuprates' high temperature strange metal phase. Alternatively, it has been recognized that strange metal transport originating from proximity to a quantum critical point (QCP) is inherently hydrodynamic\cite{sachdevhydro,zaanencond}, the result of electronic degrees of freedom appearing to behave quasi-classically because their dynamics scale only with the thermodynamic temperature $T$ and are dominated by electron-electron ($ee$) scattering\cite{sachdev}. Recent experiments have also suggested hydrodynamics may be responsible for certain universal aspects of transport in the cuprates\cite{mackenzie}. Here, we consider the cuprates from this alternative limit and treat their low energy electron matter as a hydrodynamic fluid. Specifically, we perform a rudimentary estimate of $\eta/s$ for optimally doped Bi$_{2}$Sr$_{2}$CaCu$_{2}$O$_{8+\delta}$ from a kinematic perspective.  While not a true measure of the viscosity,  this viscosity-like parameter indicates the strongly interacting cuprate electron fluid is essentially a perfect liquid along with the QGP and the UFG, approaching the holographic bound originally proposed by Kovtun, Son and Starinets (KSS) using the Anti-de Sitter Space/Conformal Field Theory (AdS/CFT) correspondence\cite{kss}

\begin{equation}\label{etas1}
    \frac{\eta}{s}\geq\frac{\hbar}{4\pi k_{B}}.
\end{equation}

Characterizing the total electrical conduction by viscosity represents a departure from our usual conception of transport in solids. Hydrodynamics in the presence of an ionic lattice requires momentum and energy to be locally conserved by the electron fluid interacting primarily with itself and dissipating disturbances collectively only at much later times\cite{kivelsonviscosity}. Mathematically this requirement is expressed as $\hbar/\tau_{ee}\gg\hbar/\tau_{e-lat}$, where $e-lat$ denotes electron-lattice interactions and $\tau$ is a scattering time; its defeat is a near-universal feature of transport in solids leading, for instance, to high-$T$ resistivity saturation at the Ioffe-Regal limit. The dominance of phonon and other Umklapp processes over pure $ee$ processes usually short-circuits true hydrodynamic flow even at high temperatures. However $\hbar/\tau_{ee}\gg\hbar/\tau_{e-lat}$ has been verified directly in Bi2212 by, for example, time resolved ARPES\cite{perfetti2ppe,rameautrarpes}. Further, the transport scattering rate $\hbar/\tau_{tr}$ is known to be dominated by electronic interactions rather than phonons or impurities. This behavior is a natural outgrowth of Bi2212's doping from a parent Mott insulator in which the $ee$ interactions rule \emph{a priori}. Further, hydrodynamic transport provides a plausible mechanism for the violation of Ioffe-Regal limit\cite{levinprl}. While the precise mechanism by which long time scale viscous dissipation leads to a finite electrical resistivity is not known, several possibilities (beyond the scope of this work) have been suggested\cite{kivelsonviscosity,zaanencond,sachdevhydro,hartnoll3}.

Lacking a true ``electrical viscometer" we appeal to the semiclassical nature of the strange metal to obtain a phenomenological estimate of $\eta/s$. Though Bi2212 does not host true quasiparticles, quasiparticle-like excitations are well-enough defined at the Fermi level $E_{F}$ and Fermi momentum $k_{F}$ that $\tau$ remains meaningful and a Boltzmann description of the fluid is still possible\cite{allen,varmalong,sachdevhydro}. Historically, analysis of ARPES data from the cuprates has proceeded accordingly. Treating nodal excitations imbued with sharp Lorentzian spectral peaks characteristic of quantum lifetime processes has, for instance, enabled the observation of quantum criticality in single particle lifetimes\cite{valla2}, the explanation of bulk transport properties in terms of microscopic origins\cite{vishik} and indeed underlies the entire many-body Greens function approach to understanding the electronic structure of the cuprates\cite{greenfunbook}.

%A kinetic treatment of the classical, viscous ideal gas shows $\eta=n k_{B} T \tau_{p}$\cite{huang} exactly, where $n$ is the particle density. Because $n k_{B} T$ is the 2D kinetic energy density $\varepsilon(T)$, $\eta(T)=\varepsilon(T)\tau_{p}(T)$\cite{kss}, $\tau_{p}(T)$ being the momentum transport rate appropriate to hydrodynamics.

One approach to estimating the viscosity of a fluid is to generalize the classical result that $\eta(T)=\varepsilon(T)\tau_{p}(T)$ where $\varepsilon(T)$ is the kinetic energy density and $\tau_{p}=\tau_{tr}$ is the momentum (or transport) relaxation rate appropriate to hydrodynamics\cite{njphys,kss}. For example, the viscosity of the classical ideal gas is exactly $\varepsilon(T)\tau_{p}=nk_{B}T\tau_{p}$ where $n$ is particle density\cite{huang}. The kinetic approach to Fermi liquid theory similarly yields $\eta_{FL}\sim \varepsilon(T)\tau_{tr}(T)$ up to a constant close to unity\cite{oldvisc,klevin}. In practice $\eta_{FL}$ turns out to be rather large because it scales with the large Fermi energy intrinsic to true metals. Graphene has been predicted to host a nearly perfect fluid\cite{schmalian} in the sense of Eq. \ref{etas1} in part because it can be easily be brought into a semiclassical regime, in which case $\varepsilon(T)\sim T$. In the case of graphene, as well as topological insulators (TI's), the relevant energy scale is taken to be the Dirac point energy, $E_{D}$, rather than $E_{F}$. So long as $E_{D}-\mu\lesssim k_{B}T$ (where $\mu$ is the chemical potential) and $k_{F}\ell\ll1$ these materials remain in the classical hydrodynamic limit, where $\ell\propto\tau^{-1}$ is the electronic mean free path. Optimally doped Bi2212 on the other hand achieves the same $T$-scaling in the normal state by virtue of its proximity to a QCP\cite{sachdev,varmalong,valla2} so that $\varepsilon(T)$ is given by the thermal kinetic energy per particle and not $E_{F}$. Below $T_{c}$, as well as above it while a fluctuating superconductivity persists, the $d$-wave nature of the superconducting gap $\Delta_{SC}$ ensures the existence of a nodal point playing the same role as $E_{D}$ in graphene and TI's. That, as well as the linearity of the band in the vicinity of $\mu$, also preserves an approximate Lorentz invariance\cite{schmalian}. Because the nodal point is pinned to $\mu$, the system remains in the classical limit. Further, below $T_{c}$ viscosity only has meaning for the normal, nodal component of the system because the superfluid component has neither entropy nor viscosity. A similar situation holds for superfluid Helium in the two fluid picture.

We proceed to approximate Eq. \ref{etas1} by replacing classical expressions for $\varepsilon(T)$ and $s(T)$ in $\eta/s=\varepsilon(T)\tau_{p}(T)/s(T)$ with those respecting Fermi-Dirac (FD) statistics and separate the thermodynamic and dynamical quantities, respectively: $T_{\eta}(T)\equiv\varepsilon(T)/s(T)$ and $\tau_{p}(T)$. Formally this approach only requires knowledge of $T$, $\tau_{p}(T)$ and the renormalized single particle densities of states (DOS) $g_{T}(\omega)$ with binding energy $\omega=E-E_{F}$. Taking $g_{T}(\omega)$ from experiment captures effects due to the pseudogap, strong coupling, etc. not easily reproduced by theory. It is by this means that the ARPES spectrum readily gives access to collective properties of electrons such as order parameters and thermal distribution functions. This procedure explicitly ignores collective excitations that do not renormalize the single particle spectrum, as appropriate to $\eta$, and in considering only the kinetic energy density no further assumptions of this sort are needed anyway\cite{levinprl}. Below $T_{c}$, superconductivity itself is entirely reflected in the renormalization of $g_{T}(\omega)$.

%The kinetic approach to Fermi liquid theory similarly yields $\eta_{FL}\sim \varepsilon(T)\tau_{tr}(T)$ up to a constant close to unity\cite{oldvisc}\cite{klevin}. In those cases $\eta$ turns out to be rather large because the energy scale relevant for transport in a true metal is $E_{F}$. Graphene has been predicted to host a nearly perfect fluid\cite{schmalian} in the sense of Eq. \ref{etas1} in part because it can be easily be brought into a semiclassical regime, $T\gtrsim E_{F}$, when the Dirac point is brought close to the chemical potential, in which case the dominant energy scale for transport is, like the classical gas, $T$. Optimally doped Bi2212 achieves the same $T$-only scaling for transport in the strange metal region of the phase diagram by virtue of its proximity to a QCP\cite{sachdev}\cite{valla2}\cite{varmalong}. $\varepsilon(T)$ is thus given by the thermal kinetic energy per particle and \emph{not} the Fermi energy.

Combining the above considerations produces our quantum critical approximation to $\eta/s$:
\begin{equation}\label{etas2}
    \frac{\eta}{s}\cong T_{\eta}\tau_{p}
\end{equation}
where $\hbar/\tau_{p}$ is determined from ARPES lineshape analysis and
\begin{equation}\label{Teta}
    T_{\eta}=\frac{\int_{-\infty}^{\infty}\omega[\tilde{g}_{T}(\omega)-\tilde{g}_{0}(\omega)]d\omega}{-k_{B}\int_{-\infty}^{\infty}[\ln(f_{T})+(f_{T}^{-1}-1)\ln(1-f_{T})]\tilde{g}_{T}(\omega)d\omega}
\end{equation}
where $f_{T}=(1+e^{\frac{-\omega}{k_{B}T}})^{-1}$ is the FD distribution, $\tilde{g}_{T}(\omega)=f_{T}g_{T}(\omega)$ is proportional to the ARPES spectrum integrated over the full Brillouin zone (BZ) and $\tilde{g}_{0}(\omega) = f_{0}[\tilde{g}_{T}(\omega)+\tilde{g}_{T}(-\omega)]$ \cite{ashcroft}. Note that only spectral weight within $\sim 4 k_{B} T$ of $E_{F}$ contributes significantly to the integrals of Eq. \ref{Teta}, numerical prefactors and proportionality constants cancel and gapped portions of the Fermi surface contribute far less to Eq. \ref{Teta} than do gapless excitations about the nodes.

Realistic absolute values of $\tau_{p}(T)$ are notoriously difficult to calculate from first principals for even the simplest systems, let alone for the cuprates, for which the origin of the linear-in-$T$ scattering rate for $T>T_{c}$ remains a mystery. However because $T_{\eta}(T)$ relies only upon the DOS, it is readily calculable from a model of the low energy band structure. $T_{\eta}(T)$ can also be estimated by purely analytical means assuming only a general form for the energy dependence of the low energy DOS. Analytical and numerical evaluation of $T_{\eta}$ first from first principles and then using a simple tight binding model and the phenomenological model due to Yang, Rice and Zhang (YRZ)\cite{yrz} can be found in the Appendix.

\section{Computation of $T_{\eta}$}\label{tetaderiv}

\begin{figure}
  \includegraphics[scale = .8, bb = 11 68 587 272]{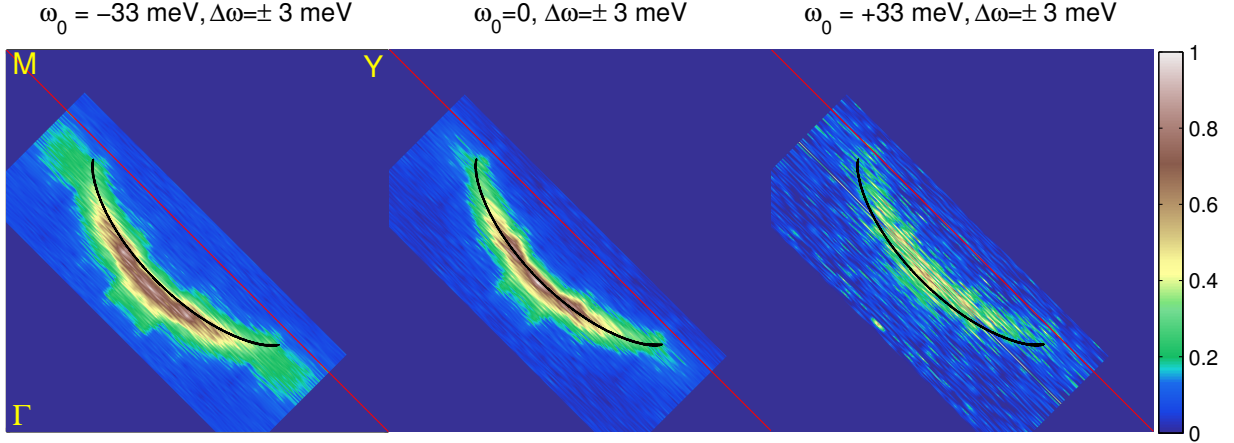}\\
  \caption{(color online) Scaled intensity maps for the Bi2212 Fermi surface at 95 K. From left to right: $\omega=-33$ meV, $\omega=0$, $\omega = +33$ meV. The integration window is $\pm3$ meV. The experimental details are described in the methods section of the main text. The red line shows the antiferromagnetic zone boundary and the black line shows the visible portion of the Fermi surface for $x=0.16$ calculated using the YRZ model as described above.}\label{mapfig}
\vspace{20pt}
\end{figure}

\begin{figure}
  \includegraphics[scale = .8, bb = 11 68 587 272]{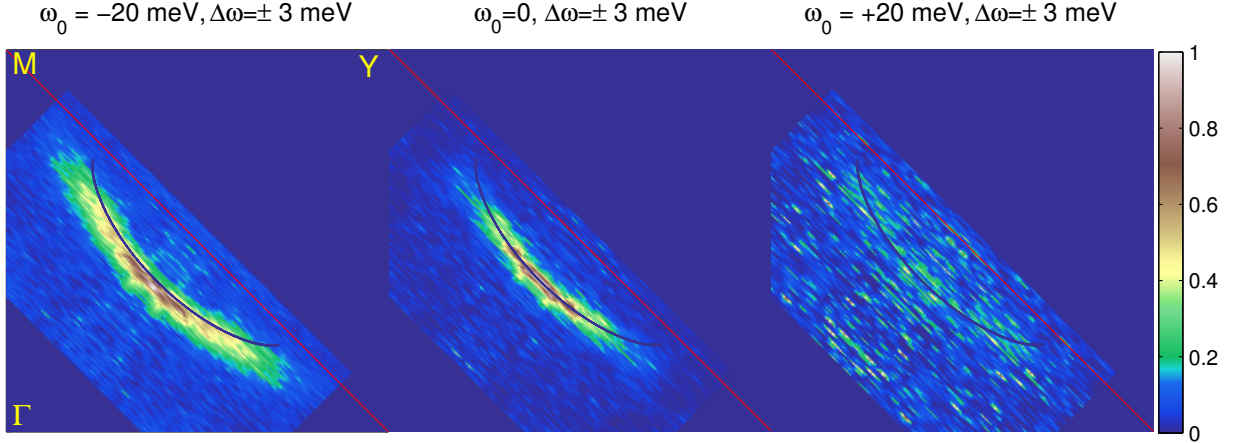}\\
  \caption{(color online) Scaled intensity maps for the Bi2212 Fermi surface at 60 K. From left to right: $\omega=-20$ meV, $\omega=0$, $\omega = +20$ meV. The integration window is $\pm3$ meV. The experimental details are described in the methods section of the main text. The red line shows the antiferromagnetic zone boundary and the black line shows the visible portion of the Fermi surface for $x=0.16$ calculated using the YRZ model as described above.}\label{mapfig2}
\end{figure}

The single particle spectral function $A(\vec{k},\omega)$, at a given temperature $T$, is related to the retarded single particle Green's function by
\begin{equation}
A(\vec{k},\omega)=-\frac{1}{\pi}|\mathrm{Im} G^{R}(\vec{k},\omega)|.
\end{equation}
where $\omega=E-E_{F}$ is the binding energy referenced to the Fermi energy $E_{F}$. In ARPES we measure a photoelectron intensity $I(\vec{k},\omega)$ (after kinematic conversion from emission angles $\theta$ and $\phi$ to momentum $\vec{k}$) proportional to $A(\vec{k},\omega)$ such that
\begin{equation}
I(\vec{k}',\omega')=\varsigma (|\mathcal{M}_{fi}(\vec{k},\omega)|^{2}f(\omega,T)A(\vec{k},\omega))\otimes R(\vec{k}-\vec{k}',\omega-\omega')
\end{equation}
where $\varsigma$ is a constant of proportionality, $|\mathcal{M}_{fi}(\vec{k},\omega)|^{2}$ is a dipole transition matrix element that in general depends on photon energy, polarization and angle of incidence as well as possible final state effects and $\otimes R(\vec{k}-\vec{k}',\omega-\omega')$ denotes convolution by a (usually Gaussian) instrumental resolution function. Resolution broadening is removed prior to other analysis by Lucy-Richardson deconvolution, as has been described extensively elsewhere\cite{jonelecspec}, so primes are dropped from here on. Since all measurements on a given sample are performed at a single photon energy we will also take its contribution to the intensity to be constant and absorb it into the overall constant of proportionality $\varsigma$. $\varsigma$ contains additional proportionalities such as photon flux, electron detector efficiency and a host of other contributions internal and external to the sample that render the measured ARPES spectrum proportional to the absolute value of the spectral function which encodes the probability for electron removal (or addition) per $\vec{k}$ and $\omega$. In the small energy range (on the order of $\pm$100 meV at the most) about $E_{F}$ we are interested in for evaluation of $U(T)/S(T)$ we shall take $|\mathcal{M}_{fi}(\vec{k},\omega)|^{2}$ to be constant in $\omega$ and remove the $\vec{k}$ dependence, which is slow in the nodal region for this photon energy and the band of interest, and normalize it to the incoherent background then absorbing it into $\varsigma$. This approximation works here because the integrals of Eq. \ref{Teta} are heavily dominated by states at and near $E_{F}$; antinodal states in the pseudogap regime do not contribute appreciably to $T_{\eta}$ either above or below $T_{c}$. Constant energy intensity maps shown in Fig. \ref{mapfig} for $T = 95$ K at $E_{F}$ and $\pm 4 k_{B} T$, respectively, illustrate this point. The effect of a rapidly changing DOS is somewhat more dramatic below $T_{c}$ as illustrated for the $T = 60$ K intensity maps, Fig. \ref{mapfig2}.

Rearranging the remaining terms we find
\begin{equation}
A(\vec{k},\omega)\simeq\frac{I(\vec{k},\omega)}{\varsigma f_{T}}
\end{equation}
where
\begin{equation}\label{dosdef}
g_{T}(\omega)=\int_{\mathrm{BZ}}A(\vec{k},\omega)d\vec{k}=\frac{1}{\varsigma f_{T}}\int_{\mathrm{BZ}}I(\vec{k},\omega)d\vec{k}.
\end{equation}
Here BZ (Brillouin Zone) denotes integration over all $\vec{k}$ in the first BZ or, by symmetry, just the irreducible eighth of the BZ symmetrized into the first quadrant. $f_{T}$ comes out of the integral because it depends only upon $\omega$. Here $\varsigma$ absorbs the actual fraction of the BZ measured, factors of $\pi$, degeneracy factors, etc.

Our goal is to use ARPES data to evaluate
\begin{equation}\label{tetaeq1}
T_{\eta}=\frac{\varepsilon(T)}{s(T)}
\end{equation}
where $\varepsilon(T)=(U(T)-U(0))/V$ is the free energy density, $s(T)=S(T)/V$ is the entropy density and, after canceling volume factors $V$, $S(T)$ is the entropy and $U(T)$ is the total thermodynamic energy. $U(0)$ is the ground state energy to which $U(T)$ is referenced. $s(T)$ and $u(T)=U(T)/V$ are in general given by the equations\cite{ashcroft,randeria2}
\begin{equation}\label{sofT}
    s(T)=-k_{B}\int_{-\infty}^{\infty}[f_{T}\ln(f_{T})+(1-f_{T})\ln(1-f_{T})]g_{T}(\omega)d\omega
\end{equation}
and
\begin{equation}\label{uofT}
    u(T)=\int_{\mathrm{BZ}}\int_{-\infty}^{\infty}(\omega+\epsilon_{\vec{k}}) f_{T} A_{T}(\vec{k},\omega) d\omega d\vec{k}
\end{equation}
respectively. In Eq. \ref{uofT} $\varepsilon_{\vec{k}}$ is the bare electron dispersion. Separating the $\omega$ and $\varepsilon_{\vec{k}}$ terms yields
\begin{equation}
u(T)=\int_{-\infty}^{\infty}\omega f_{T} g_{T}(\omega) d\omega + \int_{\mathrm{BZ}}\epsilon_{\vec{k}} [\int_{-\infty}^{\infty} f_{T} A_{T}(\vec{k},\omega) d\omega] d\vec{k}.
\end{equation}

While $g_{T}(\omega)\propto A(\vec{k},\omega)$ ARPES measures only occupied states $I(\vec{k},\omega)\propto f_{T}(\omega)A(\vec{k},\omega)$. It is therefor useful to redefine Eqs. \ref{sofT} and \ref{uofT} in terms of occupied DOS $\tilde{g}_{T}(\omega)$:
\begin{equation}\label{dos}
\tilde{g}_{T}(\omega)=f_{T}g_{T}(\omega)=\int_{\mathrm{BZ}}I(\vec{k},\omega)d\vec{k}.
\end{equation}
To calculate $U(T=0)$ we require $\tilde{g}_{0}(\omega)$ here defined by extrapolating the state at any given $T$ to $T=0$ by ``lowering the temperature" of the Fermi function
\begin{equation}\label{g0}
    \tilde{g}_{0}(\omega) = f_{0}[\tilde{g}_{T}(\omega)+\tilde{g}_{T}(-\omega)].
\end{equation}
This procedure shifts (physical) spectral weight from above to below $E_{F}$ using symmetrization (which removes the effect of the Fermi function from the spectrum) and then cuts off the spectrum at $E_{F}$ with the step function $f_{0}$. This procedure mimics the effect of going to $T=0$, effectively implementing a ``band structure" approximation. Note also that because all weight above $E_{F}$ is set to zero at the end by $f_{0}$ no unphysical weight is produced above $E_{F}$ on the unoccupied side of the spectrum and no assumption of particle-hole symmetry or asymmetry is required. Then
\begin{equation}\label{thermal}
u(T)-u(0)=\int_{-\infty}^{\infty}\omega [\tilde{g}_{T}(\omega)-\tilde{g}_{0}(\omega)] d\omega + \int_{\mathrm{BZ}}\epsilon_{\vec{k}} [\int_{-\infty}^{\infty} [f_{T} A_{T}(\vec{k},\omega)-f_{0} A_{0}(\vec{k},\omega)] d\omega] d\vec{k}
\end{equation}
where, first performing the $\omega$ integral in Eq. \ref{thermal}, the $\epsilon_{\vec{k}}$ term can be seen to go to zero by inspection because the total spectral weight in $f_{T}A(\vec{k},\omega)$ is conserved between temperatures. This is different from the case of evaluating, for example, the energy difference between normal and superconducting spectral functions. After performing the integrals in Eq. \ref{thermal} and dividing by Eq. \ref{sofT} all constants absorbed into $\varsigma$ cancel between numerator and denominator and we are left with Eq. \ref{Teta}.

\section{Experimental Methods and Results}\label{expsec}
\subsection{Measurement of $T_{\eta}$}

\begin{figure}
  \includegraphics[width = 80mm, bb = 14 196 582 552]{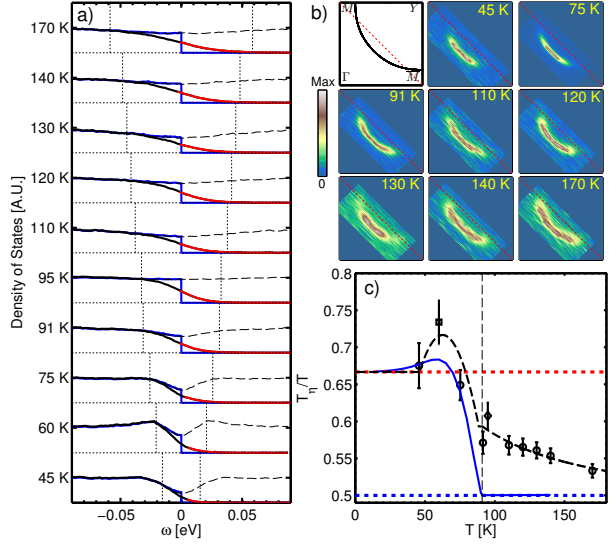}\\
  \vspace{80pt}
  \caption{(color online) Panel a) Experimental DOS for Bi2212. Raw DOS (solid black lines) are generated by integrating data over regions of the BZ demarcated by the Fermi surface maps in panel b). Raw DOS are fit above $E_{F}$ by a FD distributions (solid red lines) ensuring a smooth approach to zero intensity at high energies. $\tilde{g}_{0}(\omega)$ (solid blue lines) and symmetrized DOS (black lines) are also shown. All DOS in the figure are normalized to unity at high $(-\omega)$ and offset as indicated by the horizontal dashed black lines. Vertical dashed black lines indicate $\pm 4 k_{B}T$ for each $T$. b) Fermi surface maps corresponding to regions of the BZ measured in ARPES used to generate the DOS in panel a). Red dashed lines show the zone boundary of the underlying antiferromagnetic spin lattice. The 95 K and 60 K maps are shown in Figures \ref{mapfig} and \ref{mapfig2}, respectively. The upper-left panel shows a schematic of the underlying tight binding Bi2212 FS (black line) as described in the Appendix. c) $T$ dependence of $T_{\eta}/T$ (Eq.\ref{Teta}). (solid blue line) Theoretical $T_{\eta}/T$ for the tight binding model. (Black circles, a diamond and a square point) $T_{\eta}/T$ derived by applying Eq. \ref{Teta} to the experimental DOS in panel a). Different symbols apply to different samples. Error bars reflect the uncertainty of the chemical potential, which was 0.5 meV. Dotted red and blue lines demarcate $T_{\eta}/T$ equal to $(2/3)$ and $(1/2)$, respectively. (Black dashed line) phenomenological fit to the data used to scale ARPES scattering rates for Fig. \ref{etasfig}}\label{dosfig}
\end{figure}

Optimally doped single crystals of Bi2212 were grown using the floating zone method. $T_{c}$ was checked using SQUID magnetometery. The ARPES experiments were carried out at beamline U13UB of the National Synchrotron Light Source. Samples were mounted with the entrance slit of the hemispherical electron spectrometer along the Bi2212 $\Gamma-Y$ direction and cleaved \textit{in situ} at the lowest measured $T$ for each sample at the chamber base pressure of $8\times10^{-11}$ Torr. The chemical potential was referenced for each sample to a gold wire in electrical contact with the Bi2212 samples. $T$ was measured using a silicon diode mounted close to the samples. The temperature was ramped at a rate of 0.5 K/minute to prevent outgassing and minimize mechanical stress on the samples between sweeps of the Brillouin zone. The photon energy was set to 16.5 eV for all measurements and was polarized along the $M-\overline{M}$ plane. The matrix elements associated with this photon energy and relative polarization allow the observation of only the Bi2212 bonding band. Spectra were recorded using a Scienta SES-2002 hemispherical electron spectrometer. The total instrumental resolution (beamline + spectrometer) was set to 12.5 meV (Gaussian full width at half maximum) and angular resolution of $0.1^{\circ}$. These parameters were used as input for the Lucy-Richardson (LR) algorithm used to deconvolve instrumental broadening from the raw data\cite{jonelecspec}. The LR algorithm was set to run for three iterations on all 2D spectra. DOS were produced by trapezoidal integration across the $k_{x}$ and $k_{y}$ directions of the 3D data sets produced at each $T$. Sample surface quality and orientation was checked after the end of each run using low energy electron diffraction (LEED). For the sample on which many $T_{\eta}$ were recorded in a single run (marked by circles in Fig. \ref{dosfig}) $T$ was first raised and then lowered. The chronological order temperatures were recorded was 75 K, 91.5 K, 120 K, 140 K, 170 K, 130 K, 110 K, 45 K acquired over three days of continuous collection; both the ARPES and subsequent LEED showed minimal sample aging over this period.

In Fig. \ref{dosfig}a) we show $T$-dependent DOS acquired on several samples by integrating ARPES spectra over the regions of the BZ delineated by the intensity maps at $E_{F}$ shown in Fig. \ref{dosfig}b). The result of applying Eq. \ref{Teta} to experimental DOS is shown in Fig. \ref{dosfig}c along with theoretical $T_{\eta}(T)/T$ for a simple tight binding model of Bi2212. These results for $T_{\eta}/T$ are well understood analytically using appropriate energy dependent DOS $g_{T}(\omega)\propto\omega^{\alpha}$ with $\alpha>-1$. The analytical approximation to Eq. \ref{Teta},
\begin{equation}\label{shimul1}
    T_{\eta} \simeq\left( {\frac{{\alpha  + 1}}{{\alpha  + 2}}} \right)T
\end{equation}
is derived explicitly from the thermodynamic grand potential in the Appendix. An $\omega$-linear dispersion through $E_{F}$, as often occurs in real 2D systems, has an $\omega$-independent DOS near $E_{F}$ with $\alpha=0$ giving $T_{\eta}/T=(1/2)$.\cite{ashcroft} A Dirac cone-like dispersion, such as occurs in the nodal region of Bi2212 for $T\lesssim T_{c}$, as well as for heavily underdoped samples about the nodes for $T_{c}<T<T^{*}$ ($T^{*}$ the pseudogap temperature) gives $g_{T}(\omega)\propto\omega$, with $\alpha=1$, yielding $T_{\eta}/T=(2/3)$. Deviations of $g_{T}(\omega)$ from a simple power law result in more complicated behavior. Nevertheless, Fig. \ref{dosfig}c indicates that despite the presence of strong interactions and a relatively small pseudogap, Eq. \ref{shimul1} is reasonably accurate.

\begin{figure}
  \includegraphics[scale=.8, bb = 32 71 556 323]{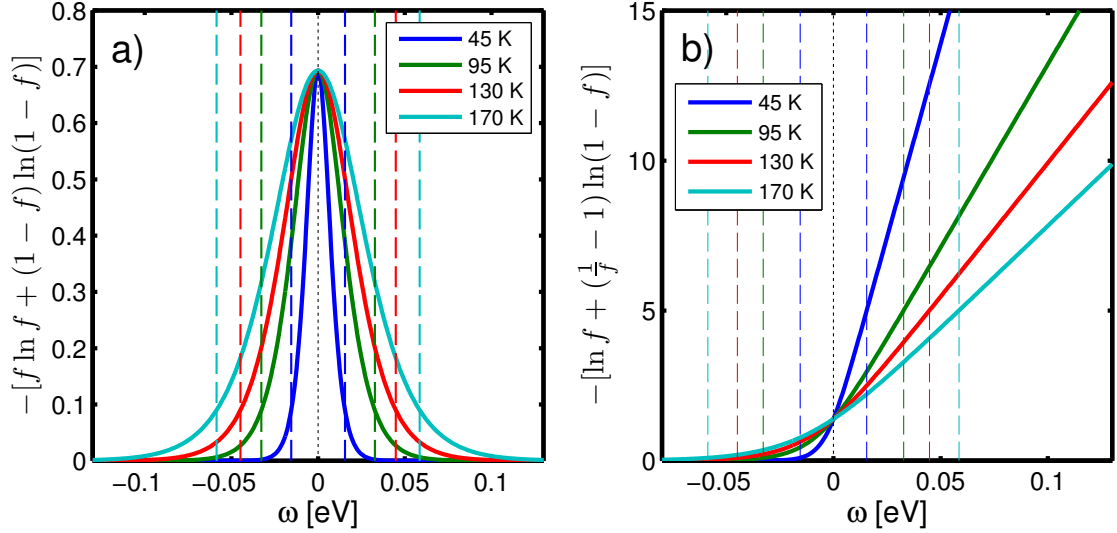}\\
  \caption{(color online) Panel a) shows the Gaussian distribution that weights the full DOS when calculating $s(T)$. Panel b) shows the factor multiplying the experimental ``occupied states" DOS actually measured directly in ARPES. In both panels the vertical dotted lines denote $\pm 4 k_{B} T$.}\label{sfactor}
\end{figure}

In practice, extraction of these results from the data as $T$ is lowered is not trivial. Any intensity noise in the measured $g_{T}(\omega)$ appearing at high energies in the ARPES spectrum can cause unphysical or misleading results when evaluating Eq. \ref{Teta}. The reason for this can be deduced from examination of the factors in Eq.'s \ref{sofT}, \ref{uofT} and \ref{Teta} that weight the measured (occupied states) DOS $\tilde{g}_{T}(\omega)$ and the ``full" DOS $g_{T}(\omega)$, respectively. The Fermi factors weighting the full and occupied DOS in Eq. \ref{sofT} are plotted for several temperatures in Fig. \ref{sfactor}. The entropy weighting factor used when considering a full DOS, as in a band structure calculation, is essentially a Gaussian distribution centered at $E_{F}$ and extending to $\sim\pm 4 k_{B} T$ above and below $E_{F}$. On the other hand, removal of a factor of $f_{T}(\omega)$ into the measured, occupied DOS in the denominator Eq. \ref{Teta} has the effect of causing the spectral weight below $E_{F}$ to be weighted somewhat less relative to the full Gaussian, and spectral weight above $E_{F}$ to contribute increasingly. In fact, the weight above $E_{F}$, which decreases exponentially due to the Fermi cutoff, has a linear in $\omega$ increase in weighting. Since in practice ARPES does not detect anything much more than $\sim 4 k_{B} T$ above $E_{F}$ \cite{yang2} contributions to the integrals Eq. \ref{sofT} or \ref{Teta} increase exponentially with $\omega$.

\begin{figure}
  \includegraphics[scale=.8, bb = 7 67 563 334]{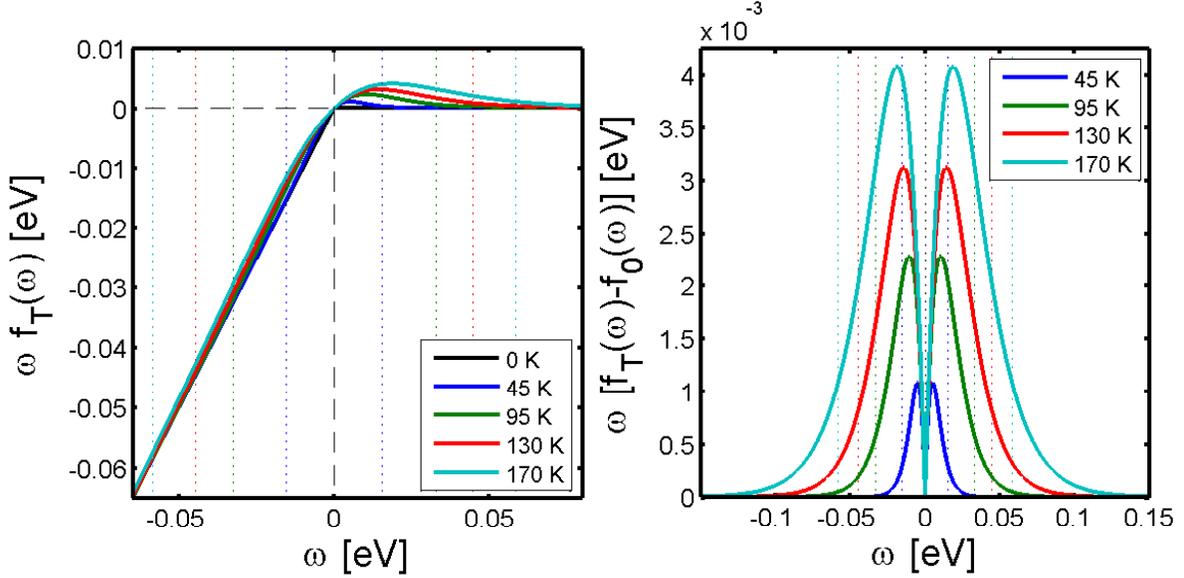}\\
  \caption{(color online) Panel a) shows the weighting factor for $u(T)$, $\omega f_{T}(\omega)$ including at $T=0$. Panel b) shows $\omega(f_{T}(\omega)-f_{0}(\omega))$. Vertical lines in both panels denote $\pm 4 k_{B} T$.}\label{ufactor}
\end{figure}

A similar effect occurs for the weighting factor $\omega f_{T}(\omega)$ in Eq. \ref{uofT}. This factor is plotted for several temperatures in panel a) of Fig. \ref{ufactor}. In panel b) we plot $\omega(f_{T}(\omega)-f_{0}(\omega))$ for the same temperatures. While this is not strictly physical because $f_{T}(\omega)$ and $f_{0}$ weight $g_{T}(\omega)$ and $g_{0}(\omega)$, respectively, it demonstrates the relative importance of excitations above $E_{F}$. In fact, it is because $u(T)$ goes to $0$ at $E_{F}$ while $s(T)$ is maximal at $E_{F}$ for a given $T$ that $T_{\eta}$ is so sensitive to the opening of a gap around the Fermi surface.

\begin{figure}
  \includegraphics[scale = .8, bb = 0 0 545 288]{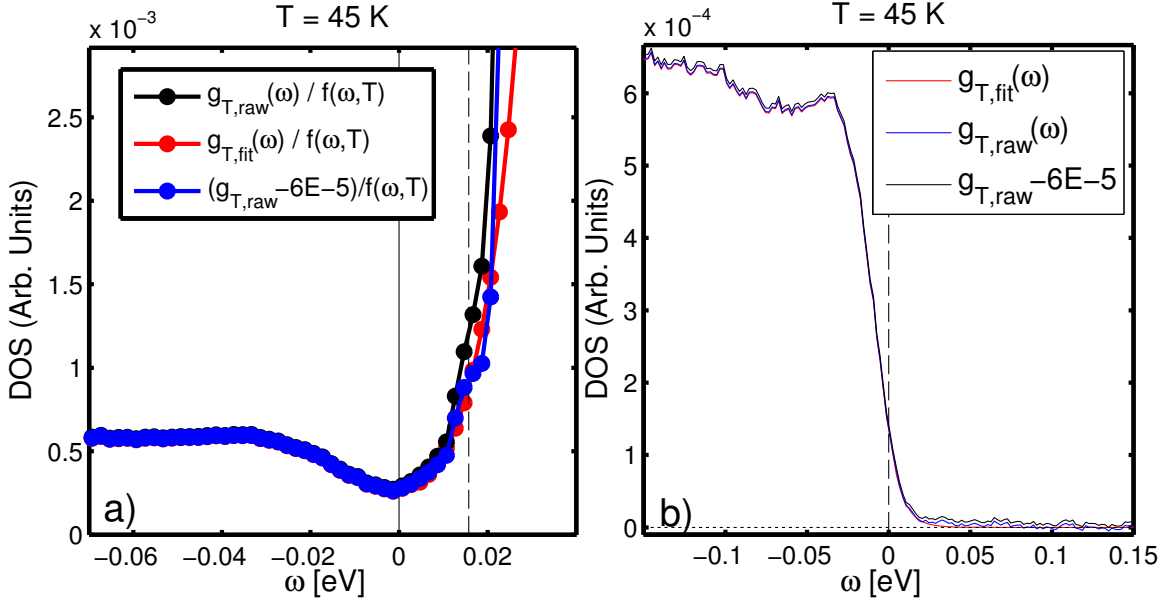}\\
  \caption{(color online) a) (black) raw DOS, $g_{T}(\omega)$, divided by $f(T,\omega)$ after subtracting just the minimum intensity value, (red) DOS fit and replaced at high $\omega$ by a FD function, background subtracted to fit zero, then divided by the FD function and (blue) raw DOS with an average background subtraction of $6\times10^{-5}$ before FD division. b) DOS before FD division (red) with fit and subtracted, (blue) minimum value subtracted and (black) minimum value + $6\times10^{-5}$ subtracted.}\label{divide}
\end{figure}

%\begin{figure}
%\vspace{200pt}
%  \includegraphics[scale=.8, bb = 1 7 340 543]{residuals.eps}\\
%  \caption{(color online) Plots of $\delta I=I_{fit}-I_{measured}$ from Fig. 1c or the main text. Here, $I_{fit}$ refers to the red solid lines in Fig. 1c of the main text and $I_{measured}$ refers to the data, which was integrated over $k_{x}$ and $k_{y}$ and scaled to unity at high negative $\omega$.}\label{residuals}
%\end{figure}
The experimental problem faced here amounts to dividing out the FD distribution from the data without allowing the exponential blow-up of noise far above $E_{F}$ that commonly occurs during this procedure to effect the extraction of $u(T)$ and $s(T)$. What's more, such noise can make it difficult to locate the true ``zero" level of $\vec{k}$ integrated data; the removal of such background and smooth zeroing of data above $E_{F}$ are vital to the successful evaluation of Eq. \ref{Teta}. Some possibilities for handling this are to impose a cutoff in positive $\omega$ that varies from spectrum to spectrum with e.g. statistical quality of the data, imposing a uniform and possibly arbitrary cutoff in $\omega$ across all data and working with purely symmetrized data, which imposes a possibly false particle-hole symmetry on the full DOS. As a compromise we have employed a method of fitting a FD distribution function to the high energy tails of the DOS integrated in $k_{x}$ and $k_{y}$ and then replacing the measured DOS at those $\omega$ with the fit. The advantage of this method is that the fits invariably smoothly approach zero intensity in a noise-free fashion far above $E_{F}$ so that the overall background of the spectrum can be extracted with certainty before applying Eq. \ref{Teta} to the data. The fits to the data are shown in Fig. \ref{dosfig}a of the main text as red lines overlaying the data. The maximum difference between the fits and the data they replace is on the order of $2\%$. Another measure of the efficacy of this procedure is to divide out the FD distribution from the raw $g_{T,raw}(\omega)$ with the background subtraction at high $\omega$ performed using just the minimum intensity value (to avoid negative intensities) and compare this to $g_{T,fit}(\omega)$ where we have performed the fitting procedure described above. The comparison is shown in Fig. \ref{divide}a for the $45 \mathrm{K}$ data point. The agreement is very good up to $4 k_{B} T$ but without the ambiguity of noise.

\subsection{Measurement of $\hbar/\tau_{p}$}

\begin{figure}
  \includegraphics[scale = .5, bb = 38 23 574 472]{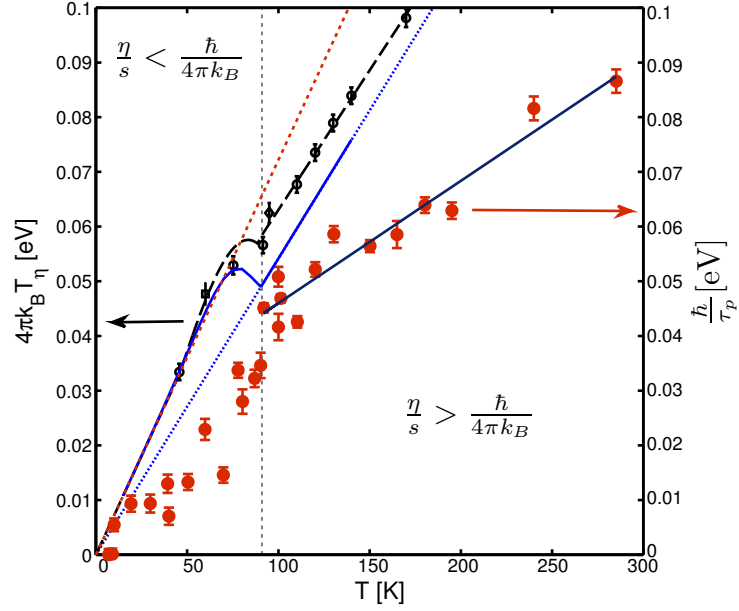}\\
  \caption{(color online) Temperature dependence of $\hbar/\tau_{p}$ derived from nodal ARPES data\cite{finedetails} for Bi2212 ($T_{c}=91$ K) (red circles). Original data\cite{finedetails}, in which inverse mean free paths $\Delta k=\ell^{-1}$ are measured directly, has been rescaled into scattering rates using the temperature-dependent Fermi velocities $v_{F}$ (in units of [eV\AA]) of Ref. [\cite*{plumb}] such that $\hbar/\tau_{\vec{k}}=v_{F}\Delta k$. The final scattering rate (red circles) is $\hbar/\tau_{p}=\hbar/\tau_{\vec{k}}-\hbar/\tau_{0}$ where for simplicity we take $\hbar/\tau_{0}=min[\hbar/\tau_{\vec{k}}(T\rightarrow 0)]=17.7$ meV. $4\pi k_{B}T_{\eta}$ lines are also plotted as a reference; the closer $\hbar/\tau_{p}$ approaches $T_{\eta}(T)$ from below, the closer $\eta/s$ is to the holographic bound after scaling by Eq. \ref{etas2}. Relevant ideal bounds consistent with Eq. \ref{etas1} include $T_{\eta}=(1/2)T$ (dotted blue line) and $T_{\eta}=(2/3)T$ (dotted red line) from Eq. \ref{shimul1}, yielding $\hbar/\tau_{p}=2\pi k_{B} T$ and $\hbar/\tau_{p}=(8/3)\pi k_{B} T$, respectively,  as well as the $T_{\eta}$ including d-wave superconductivity in the simple tight binding model (solid blue line) and $T_{\eta}$ from the present experiment on Bi2212 (black circles). The phenomenological fit to the experimental $T_{\eta}$ values is shown as the dashed black line. The navy line is a linear fit to $\hbar/\tau_{p}(T>T_{c})$.}\label{gammafig}
\end{figure}

Considering dynamics, the momentum transport rate $\tau_{p}$ entering $\eta$ is a two-particle time whereas $\tau_{k}$ measured in ARPES is single particle. While \emph{in general} one cannot extract $\tau_{p}$ from $\tau_{k}$, it is in fact possible for many 2D materials, Bi2212 included. This is because the remnant $T$- and $\omega$-independent impurity contribution to the ARPES spectral width, isolated by going to $T\approx0$, is typically more than two orders of magnitude greater than the equivalent, small contribution seen in transport\cite{hartnollimpurity,dassarma}. This occurs in Bi2212 because $\hbar/\tau_{k}$ is dominated by forward scattering induced by strong out-of-plane disorder and $\hbar/\tau_{tr}\cong\hbar/\tau_{p}$ probes only the much smaller in-plane, back scattering contribution\cite{greenfunbook,dassarma}. Once the impurity contribution is effectively removed by going to low $T$, $\hbar/\tau_{k}\cong\hbar/\tau_{tr}$ for $T>0$ because they are observed to have the same $T$-linear change in scattering rate per Kelvin. This trend is widespread\cite{tasetwo,perfettititetwo,valla2,mackenzie} and has been long appreciated in connection with the Marginal Fermi Liquid phenomenology of the cuprates\cite{varmalong}. In Eq. \ref{etas2} we therefor apply
\begin{equation}\label{lifetime}
    \frac{\hbar}{\tau_{p}(T)}\cong\frac{\hbar}{\tau_{k}(T)}-\frac{\hbar}{\tau_{0}},
\end{equation}
where $\hbar/\tau_{0}=\hbar/\tau_{k}(T=0)$ and $\hbar/\tau_{k}$ is the full width at half maximum of the Lorentzian spectral line shape at $E_{F}$ and $k_{F}$ and the forward scattering most apparent in ARPES does not dissipate electron momentum. Eq. \ref{lifetime} allows us to exploit the additional advantage of ARPES, over transport, of access to $\tau_{p}(T<T_{c})$ for the gapless nodal states of Bi2212 which were previously measured\cite{finedetails}. In Fig. \ref{gammafig} we plot $\hbar/\tau_{p}(T)$ above and below $T_{c}$. As a reference, limits on the scattering rate derived from Eq. \ref{etas1} , $\hbar/\tau_{p}\leq4\pi k_{B} T_{\eta}$, are plotted. Comparison of $\hbar/\tau_{p}(T)$ derived from ARPES data using the above procedure to that acquired using optical conductivity (in the DC limit) on similar samples by Hwang et al.\cite{opticond} indicates good agreement at $T_{c}$ with deviations on the order of ten percent as $T$ is increased towards room temperature.

\section{Discussion}\label{conc}

\begin{figure}
  %\includegraphics[scale = .5, bb = 27 31 527 517]{etasfig.eps}\\
  \includegraphics[scale = .5, bb = 27 31 527 517]{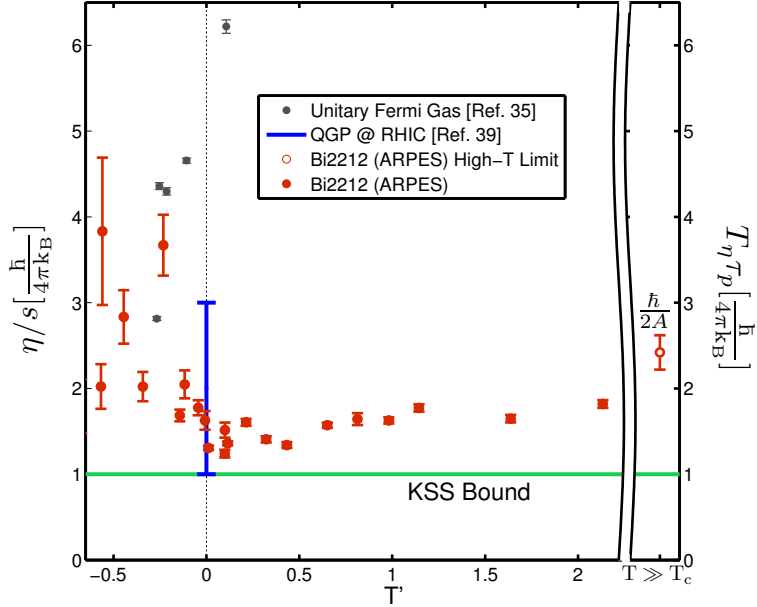}\\
  \caption{(color online) $T_{\eta}\tau_{p}$ in units of the KSS bound, $\hbar/(4\pi k_{B})$, versus $T'=(T-T_{c})/T_{c}$ evaluated using Eq. \ref{etas2} (right axis) and $\eta/s$ for the QGP and UFG (left axis). (filled red circles) ARPES for Bi2212 scaled by the phenomenological fit to $T_{\eta}$ (black dashed line in Fig. \ref{dosfig}. $T_{\eta}\tau_{p}$'s at lower $T'$ are too small to provide a reliable result. (blue bar) The range of $\eta/s$ consistent with RHIC data on the QGP at $T_{c}$ from Au+Au collisions is encompassed by the blue vertical band\cite{prcqgp}. (gray points) data from the UFG\cite{UFG1}. (open red circle) The point at $T\gg T_{c}$ is the extrapolated high $T$ value for $\eta/s\sim\hbar/2A$ of Bi2212. The KSS bound is marked by the solid green line and $T_{c}$ for Bi2212 by the vertical black dotted line.}\label{etasfig}
\end{figure}

In Fig. \ref{etasfig} we present our main findings, a plot of $\eta/s$, evaluated using Eq. \ref{etas2}, for Bi2212 versus a reduced temperature $T'=(T-T_{c})/T_{c}$. While calculations of $\eta$ for the Fermi liquid\cite{oldvisc,pines} suggest our own analysis is correct up to a factor of order one. In fact, since in 2D equipartition guarantees a prefactor not greater than unity, our results represent an upper bound to $\eta/s$ even allowing for a fully quantum calculation. We therefor find that, as defined, Bi2212 nearly saturates the holographic bound, Eq. \ref{etas1}, around $T_{c}$, showing the electronic subsystem hosted in its CuO$_{2}$ planes is a nearly perfect fluid. Immediately below $T_{c}$ the combination of a decreasing $\hbar/\tau_{p}$ and increasing $T_{\eta}(T)$ conspire to rapidly raise $\eta/s$ as $T$ is lowered. The minimum in $\eta/s$ resembles what is expected for a gas-liquid phase transition as $T$ is lowered through $T_{c}$ and is consistent with expectations of a ``check mark" shape for $\eta(T)/s(T)$ found in other strongly interacting quantum fluids\cite{njphys,sphenix}. Above $T_{c}$ we fit $\hbar/\tau_{p}(T) = AT+B$, where $A$ and $B$ are constants. At asymptotically high $T$, $AT\gg B$ and $\eta/s$ approaches a constant value, $(\eta/s)_{HT}\approx\hbar/2A$, where $A$ is the scattering rate per Kelvin and $T_{\eta}=T/2$ is assumed. The linear fit of $\hbar/\tau_{p}(T>T_{c})$, Fig. \ref{gammafig}, yields a high-$T$ estimate of $(\eta/s)_{HT}=2.42\pm0.20 [\hbar/4\pi k_{B}]$. This is quantitatively similar to the value of $\hbar/A$ extracted from transport measurements on many strongly correlated materials, including Bi2212, in Ref. [\cite{mackenzie}], supporting the hypothesis that this quantity is related to hydrodynamic transport for some materials. Further, noting $s(T_{c})\sim n k_{B}$ with $n=x$ holes/CuO$_{2}$ plane, $\eta(T_{c})\approx x\hbar/(4\pi)$, conforming to the expectation $\eta\sim n\hbar$.\cite{UFG1,Tallon1} While it might at first appear odd that a single particle measurement could yield a result so close to predictions for what is properly a many-body property, we note the equivalence of momentum and energy transport in the cuprates has long been known phenomenologically\cite{varmalong}.

 %and, presently, is ostensibly the result of strong correlations making the ARPES photohole a ``liquid of one" as it probes the Hilbert space of the remaining electron fluid that ultimately annihilates it.

While our method of evaluating $\eta/s$ is necessarily approximate it is able to fully exploit the ability to accurately measure and control the equilibrium system temperature of a system inherent to experiments in condensed matter. The striking similarity between the magnitude and $T$-dependence of $\eta/s$ surmised for the QGP\cite{njphys,glasma,sphenix} and the related quantity for Bi2212, which is an upper bound on $\eta/s$, deduced in the present work not only raises a number of questions fundamental to strongly correlated matter but also offers to illuminate our understanding of these more exotic creations. While the derivation of the viscosity bound from AdS/CFT proceeded specifically to account for the small $\eta/s$ of the QGP its application to problems in strongly interacting condensed matter is still in its infancy so far as experiment is concerned. The very existence of a holographic bound on $\eta/s$ was justified by its compatibility with the uncertainty principle, invoking energy density arguments similar to those made above\cite{kss}, with similar justifications made in the case of graphene\cite{schmalian}. In condensed matter, such considerations are encoded in the supposition that quantum critical materials like optimally doped Bi2212 obey an expression dimensionally equivalent to Eq. \ref{etas1}, and indeed nearly identical to Eq. \ref{etas2}, $T\tau_{\phi}\geq \mathcal{C} \hbar/k_{B}$, where $\mathcal{C}$ is an unspecified universal constant of order one and $\tau_{\phi}$ is a relaxation time - single or many-particle - intrinsic to the system\cite{sachdev}. The implication of our present work is therefore that $\mathcal{C}$ itself obtains a universal lower limit approximately the same as that of the KSS bound, and for the same reasons\cite{zaanenplanck}.

\clearpage
\appendix
%\section{}
\section{Analytical Evaluation of $T_{\eta}(T)$}\label{tetaanalytic}

The thermodynamic grand potential $\Phi$, for an ensemble of non-interacting fermions with dispersion $\epsilon_{\vec{k}}$ and DOS $g_{T}(\omega)$, is given by the expression\cite{schwabl},

\begin{equation}
\begin{aligned}
 \Phi  &=  - \frac{{2V}}{\beta }\int {\frac{{{d^d}k}}{{{{(2\pi )}^d}}}\ln \left( {1 + {e^{ - \beta ({\epsilon_{\vec{k}}} - \mu )}}} \right)}  \\
  &=  - \frac{{2V}}{{\beta }}\int_0^\infty  {d\omega g_{T}(\omega))\ln \left( {1 + {e^{ - \beta (\omega  - \mu )}}} \right)}
 \end{aligned}
\end{equation}
where $\beta=(k_{B} T)^{-1}$, $\mu$ is the chemical potential and $d$ is the spatial dimensionality of the system. It follows that the entropy $S$ can be computed as,
\begin{equation}
S =  - {\left( {\frac{{\partial \Phi }}{{\partial T}}} \right)_{V,\mu }}
\end{equation}
and the internal energy $U$ can be determined from the integral
\begin{equation}
U = 2V\int_0^\infty  {d\epsilon \frac{{\omega N(\omega)}}{{{e^{\beta (\omega - \mu )}} + 1}}}.
\end{equation}
Let us assume that the DOS takes on a power-law form
\begin{equation}
g_{T}(\omega)=g_{T}^{0} \omega^{\alpha}
\end{equation}
where $g_{T}^{0}$ is a constant. The grand potential can be further reduced by integrating by parts,
\begin{equation}
\begin{aligned}
 \Phi  &=  - \frac{{2V}}{\beta }\left[ {\left. {{{\left( {\int {d\omega' g_{T}(\omega' )} } \right)}_{\omega'  = \omega}}\ln \left( {{e^{ - \beta ( \omega - \mu )}} + 1} \right)} \right|_0^\infty  + \beta \int_0^\infty  {d  \omega\frac{{{{\left( {\int {d\omega' g_{T}(\omega')} } \right)}_{\omega'  =  \omega}}{e^{ - \beta ( \omega - \mu )}}}}{{{e^{ - \beta ( \omega - \mu )}} + 1}}} } \right] \\
  &=  - 2V\int_0^\infty  {d \omega\frac{{{{\left( {\int {d\omega' g_{T}(\omega')} } \right)}_{\omega'  =  \omega}}}}{{{e^{\beta ( \omega - \mu )}} + 1}}}  \\
 \end{aligned}
\end{equation}
which is simplified by the first term vanishing only if $\alpha>-1$.
Then $\Phi$ becomes,
\begin{equation}
\begin{aligned}
 \Phi  &=  - \left( {\frac{1}{{\alpha  + 1}}} \right)  \left( {2V{N_0}} \right)  \int_0^\infty  {d\varepsilon \frac{{{\epsilon ^{\alpha  + 1}}}}{{1 + {e^{  \beta (\epsilon  - \mu )}}}}}  \\
  &=  - \left( {\frac{1}{{\alpha  + 1}}} \right)\left( {2V{N_0}} \right){T^{\alpha  + 2}}{\zeta _{\alpha  + 2}}(z)
 \end{aligned}
\end{equation}
where $z$ is the fugacity defined as $z=e^{\mu/k_{B} T}$ and $\zeta_\nu(z)$ is the Fermi-Dirac integral defined as
\begin{equation}
{\zeta _\nu }(z) = \frac{1}{{\Gamma (\nu )}}\int_{0}^{\infty}  {dx\,\frac{{{x^{\nu  - 1}}}}{{{e^x}/z + 1}}}  = \sum\limits_{k = 1}^\infty  {\frac{{{{( - 1)}^{k + 1}}{z^k}}}{{{k^\nu }}}}
\end{equation}
with Gamma function $\Gamma(x)$.
For a homogenous system,  the internal energy $U$ is proportional to $\Phi$, namely
\begin{equation}
\Phi = -PV =- \frac{1}{\alpha+1} U
\end{equation}

Subsequently, the entropy can be determined from the derivative,
\begin{equation}
 S = \left( {\frac{1}{{\alpha  + 1}}} \right)   \left( {2V{g_{T}^{0}}} \right)        \left[ {(\alpha  + 2){T^{\alpha  + 1}}{\zeta _{\alpha  + 2}}(z) + {T^{\alpha  + 2}}\frac{{\partial {\zeta _{\alpha  + 2}}(z)}}{{\partial T}}} \right]
\end{equation}
and the exact expression for U/S becomes,
\begin{equation}
\frac{U}{S} = \frac{{(\alpha  + 1){T^{\alpha  + 2}}{\zeta _{\alpha  + 2}}(z)}}{{(\alpha  + 2){T^{\alpha  + 1}}{\zeta _{\alpha  + 2}}(z) + {T^{\alpha  + 2}}\frac{{\partial {\zeta _{\alpha  + 2}}(z)}}{{\partial T}}}}
\end{equation}

%However, it is a good approximation that $\mu (T) \sim a T$, over a wide temperature range,

It should be noted that $\zeta_\nu(z)$  is constant for small  and large $z$, which is not exactly the same as large or small temperatures since the chemical potential does depend on temperature. However $\frac{d\zeta_{\nu}(z)}{dT}$ is small and can be neglected in the temperature dependence of $\zeta_\nu(z)$.  Therefor, the leading order contribution to $U/S=T_{\eta}$ is
\begin{equation}\label{shimulformula}
T_{\eta} \simeq\left( {\frac{{\alpha  + 1}}{{\alpha  + 2}}} \right)T.
\end{equation}

\section{Numerical Evaluation of $T_{\eta}$}\label{tetanumerical}

For a numerical evaluation of $T_{\eta}$ we consider two forms for $A(\vec{k},\omega)$: a tight binding dispersion and the YRZ model. In both cases we introduce a d-wave superconducting gap with amplitude $\Delta_{SC}(T)=\Delta_{SC}[1-(T/T_{c})^2]$ below $T_{c}$.
In both cases $\mu$ is fixed by the required particle number at $T=0$ and approximated as constant over the temperature interval of interest.
For the comparisons in this work we neglect lifetime broadening and take $A(\vec{k},\omega)$ as a delta function.  Lifetime effects may be incorporated in a simple manner by replacing the delta function with a suitable Lorentzian: generally this broadening leads to a more slowly varying (closer to constant) $g(\omega)$.
%We also use a Lorentzian with a full width at half maximum $\Gamma = 20$meV, to take into account the observed finite lifetime of the fermions.

For the tight binding case we use
\begin{align}
A(\vec{k},\omega)&=\delta(\omega-\epsilon_{\vec{k}}), \nonumber \\
\epsilon_{\vec{k}} &= -2 t_{0} (\cos k_{x} +\cos k_{y})-4t_{0}' (\cos k_{x} \cos k_{y})-2t_{0}'' (\cos 2 k_{x} +\cos 2 k_{y}) -\mu.
\end{align}
%For the tight binding case we use
%\begin{align}
%A(k,\omega)&=\frac{\Gamma/2}{(\Gamma/2)^2+(\omega-\epsilon_k)^2}, \nonumber \\
%\epsilon_k &= -2 t_0 (\cos k_x +\cos k_y)-4t_0' (\cos k_x \cos ky)-2t_0'' (\cos 2 k_x +\cos 2 k_y) -\mu.
%\end{align}
with hopping parameters
\begin{align}
t_{0}=360 \mathrm{meV}, \quad t_{0}'= -0.3 t_{0}, \quad t'' = 0.2 t_{0}, \quad \Delta_{SC}=0.07 t_{0}.
\end{align}
Integrals over $\omega$ are then trivial, and we perform the remaining $k$-space integrals numerically.

\begin{figure}
  \includegraphics[scale = 1,bb = 0 0 391 293]{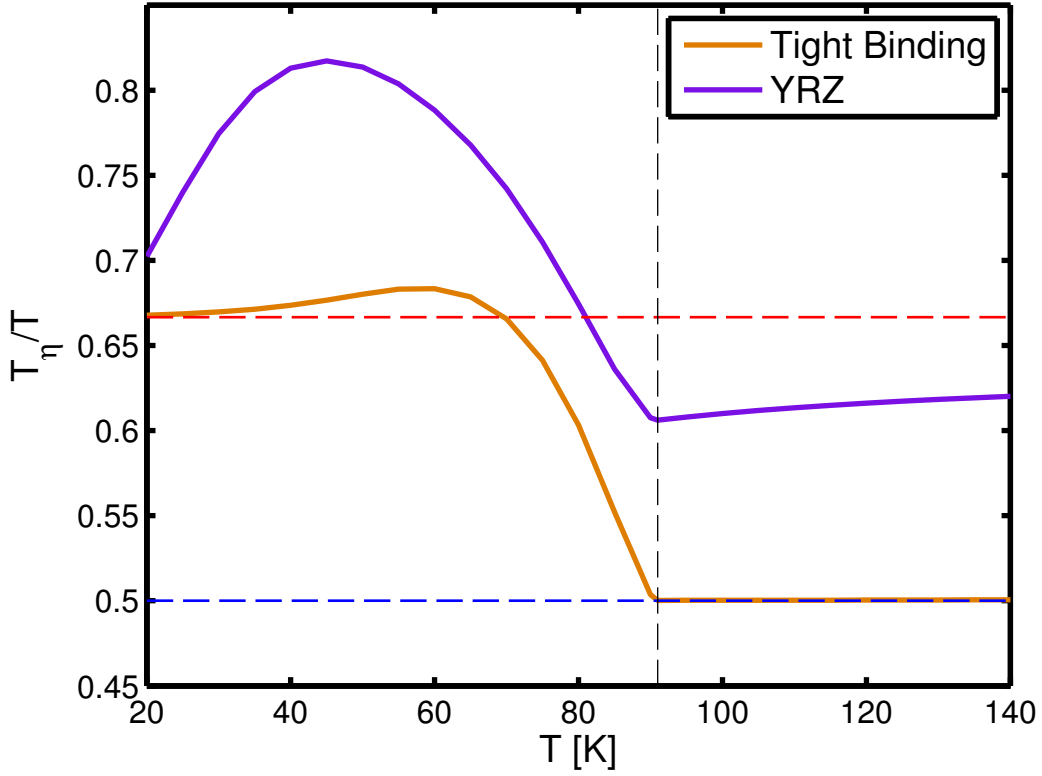}\\
  \caption{(color online) $T_{\eta}/T=U/S T$ for the tight binding (orange line) and YRZ model at $x=0.16$ (purple line), respectively. The red and blue dashed lines correspond to $T_{\eta}/T=(2/3)$ and $(1/2)$, respectively. The vertical dashed black line marks $T_{c}=91 \mathrm{K}$.}\label{andrew}
\end{figure}

For the YRZ model we use the same bare parameters as in the original formulation of the YRZ model\cite{yrz}, the only difference being that we set the pseudogap as closing at hole doping fraction $x=0.2$, higher than the critical doping $x_{c}=0.16$. Following Ref. [\cite{yrzandrewcalc}], we introduce the superconducting gap in the lower YRZ band only and again use $\Delta_{SC}=0.07 t_{0}$. $U/ST=T_{\eta}/T$ for the YRZ and tight binding models is shown in Fig. \ref{andrew}.

\begin{acknowledgments}
We would like to acknowledge helpful discussions with Alexei Tsvelik, Chris Homes, C.P. Herzog, Jan Zaanen, Tonica Valla, Philip W. Phillips, T.M. Rice, S.A. Kivelson, Peter Steinberg, Raju Venugopalan, Bj\"{o}rn Schenke and Sean Hartnoll. We'd also like to thank A.J.A. James assistance with tight binding calculations. The work at Brookhaven is supported in part by the U.S. DOE under Contract No. DE-AC02-98CH10886 and in part by the Center for Emergent Superconductivity (CES), an Energy Frontier Research Center funded by the U.S. DOE, Office of Basic Energy Sciences. S. Campbell acknowledges support from DOE grant DE-FG02-92ER40692.
\end{acknowledgments}
\clearpage


\begin{thebibliography}{41}%
\makeatletter
\providecommand \@ifxundefined [1]{%
 \@ifx{#1\undefined}
}%
\providecommand \@ifnum [1]{%
 \ifnum #1\expandafter \@firstoftwo
 \else \expandafter \@secondoftwo
 \fi
}%
\providecommand \@ifx [1]{%
 \ifx #1\expandafter \@firstoftwo
 \else \expandafter \@secondoftwo
 \fi
}%
\providecommand \natexlab [1]{#1}%
\providecommand \enquote  [1]{``#1''}%
\providecommand \bibnamefont  [1]{#1}%
\providecommand \bibfnamefont [1]{#1}%
\providecommand \citenamefont [1]{#1}%
\providecommand \href@noop [0]{\@secondoftwo}%
\providecommand \href [0]{\begingroup \@sanitize@url \@href}%
\providecommand \@href[1]{\@@startlink{#1}\@@href}%
\providecommand \@@href[1]{\endgroup#1\@@endlink}%
\providecommand \@sanitize@url [0]{\catcode `\\12\catcode `\$12\catcode
  `\&12\catcode `\#12\catcode `\^12\catcode `\_12\catcode `\%12\relax}%
\providecommand \@@startlink[1]{}%
\providecommand \@@endlink[0]{}%
\providecommand \url  [0]{\begingroup\@sanitize@url \@url }%
\providecommand \@url [1]{\endgroup\@href {#1}{\urlprefix }}%
\providecommand \urlprefix  [0]{URL }%
\providecommand \Eprint [0]{\href }%
\providecommand \doibase [0]{http://dx.doi.org/}%
\providecommand \selectlanguage [0]{\@gobble}%
\providecommand \bibinfo  [0]{\@secondoftwo}%
\providecommand \bibfield  [0]{\@secondoftwo}%
\providecommand \translation [1]{[#1]}%
\providecommand \BibitemOpen [0]{}%
\providecommand \bibitemStop [0]{}%
\providecommand \bibitemNoStop [0]{.\EOS\space}%
\providecommand \EOS [0]{\spacefactor3000\relax}%
\providecommand \BibitemShut  [1]{\csname bibitem#1\endcsname}%
\let\auto@bib@innerbib\@empty
%</preamble>
\bibitem [{\citenamefont {Andreev}\ \emph {et~al.}(2011)\citenamefont
  {Andreev}, \citenamefont {Kivelson},\ and\ \citenamefont
  {Spivak}}]{kivelsonviscosity}%
  \BibitemOpen
  \bibfield  {author} {\bibinfo {author} {\bibfnamefont {A.~V.}\ \bibnamefont
  {Andreev}}, \bibinfo {author} {\bibfnamefont {S.~A.}\ \bibnamefont
  {Kivelson}}, \ and\ \bibinfo {author} {\bibfnamefont {B.}~\bibnamefont
  {Spivak}},\ }\href {\doibase 10.1103/PhysRevLett.106.256804} {\bibfield
  {journal} {\bibinfo  {journal} {Phys. Rev. Lett.}\ }\textbf {\bibinfo
  {volume} {106}},\ \bibinfo {pages} {256804} (\bibinfo {year}
  {2011})}\BibitemShut {NoStop}%
\bibitem [{\citenamefont {Damle}\ and\ \citenamefont
  {Sachdev}(1997)}]{sachdevhydro}%
  \BibitemOpen
  \bibfield  {author} {\bibinfo {author} {\bibfnamefont {K.}~\bibnamefont
  {Damle}}\ and\ \bibinfo {author} {\bibfnamefont {S.}~\bibnamefont
  {Sachdev}},\ }\href {\doibase 10.1103/PhysRevB.56.8714} {\bibfield  {journal}
  {\bibinfo  {journal} {Phys. Rev. B}\ }\textbf {\bibinfo {volume} {56}},\
  \bibinfo {pages} {8714} (\bibinfo {year} {1997})}\BibitemShut {NoStop}%
\bibitem [{\citenamefont {Davison}\ \emph {et~al.}(2013)\citenamefont
  {Davison}, \citenamefont {Schalm},\ and\ \citenamefont
  {Zaanen}}]{zaanencond}%
  \BibitemOpen
  \bibfield  {author} {\bibinfo {author} {\bibfnamefont {R.~A.}\ \bibnamefont
  {Davison}}, \bibinfo {author} {\bibfnamefont {K.}~\bibnamefont {Schalm}}, \
  and\ \bibinfo {author} {\bibfnamefont {J.}~\bibnamefont {Zaanen}},\ }\href
  {http://arxiv.org/abs/1311.2451} {\bibfield  {journal} {\bibinfo  {journal}
  {arXiv}\ ,\ \bibinfo {pages} {arXiv:1311.2451v1}} (\bibinfo {year}
  {2013})}\BibitemShut {NoStop}%
\bibitem [{\citenamefont {Sachdev}(2001)}]{sachdev}%
  \BibitemOpen
  \bibfield  {author} {\bibinfo {author} {\bibfnamefont {S.}~\bibnamefont
  {Sachdev}},\ }\href@noop {} {\emph {\bibinfo {title} {Quantum Phase
  Transitions}}}\ (\bibinfo  {publisher} {Cambridge University Press},\
  \bibinfo {year} {2001})\BibitemShut {NoStop}%
\bibitem [{\citenamefont {Bruin}\ \emph {et~al.}(2013)\citenamefont {Bruin},
  \citenamefont {Sakai}, \citenamefont {Perry},\ and\ \citenamefont
  {Mackenzie}}]{mackenzie}%
  \BibitemOpen
  \bibfield  {author} {\bibinfo {author} {\bibfnamefont {J.~A.~N.}\
  \bibnamefont {Bruin}}, \bibinfo {author} {\bibfnamefont {H.}~\bibnamefont
  {Sakai}}, \bibinfo {author} {\bibfnamefont {R.~S.}\ \bibnamefont {Perry}}, \
  and\ \bibinfo {author} {\bibfnamefont {A.~P.}\ \bibnamefont {Mackenzie}},\
  }\href {\doibase 10.1126/science.1227612} {\bibfield  {journal} {\bibinfo
  {journal} {Science}\ }\textbf {\bibinfo {volume} {339}},\ \bibinfo {pages}
  {804} (\bibinfo {year} {2013})}\BibitemShut {NoStop}%
\bibitem [{\citenamefont {Kovtun}\ \emph {et~al.}(2005)\citenamefont {Kovtun},
  \citenamefont {Son},\ and\ \citenamefont {Starinets}}]{kss}%
  \BibitemOpen
  \bibfield  {author} {\bibinfo {author} {\bibfnamefont {P.~K.}\ \bibnamefont
  {Kovtun}}, \bibinfo {author} {\bibfnamefont {D.~T.}\ \bibnamefont {Son}}, \
  and\ \bibinfo {author} {\bibfnamefont {A.~O.}\ \bibnamefont {Starinets}},\
  }\href {\doibase 10.1103/PhysRevLett.94.111601} {\bibfield  {journal}
  {\bibinfo  {journal} {Physical Review Letters}\ }\textbf {\bibinfo {volume}
  {94}},\ \bibinfo {pages} {111601} (\bibinfo {year} {2005})}\BibitemShut
  {NoStop}%
\bibitem [{\citenamefont {Perfetti}\ \emph {et~al.}(2007)\citenamefont
  {Perfetti}, \citenamefont {Loukakos}, \citenamefont {Lisowski}, \citenamefont
  {Bovensiepen}, \citenamefont {Eisaki},\ and\ \citenamefont
  {Wolf}}]{perfetti2ppe}%
  \BibitemOpen
  \bibfield  {author} {\bibinfo {author} {\bibfnamefont {L.}~\bibnamefont
  {Perfetti}}, \bibinfo {author} {\bibfnamefont {P.~A.}\ \bibnamefont
  {Loukakos}}, \bibinfo {author} {\bibfnamefont {M.}~\bibnamefont {Lisowski}},
  \bibinfo {author} {\bibfnamefont {U.}~\bibnamefont {Bovensiepen}}, \bibinfo
  {author} {\bibfnamefont {H.}~\bibnamefont {Eisaki}}, \ and\ \bibinfo {author}
  {\bibfnamefont {M.}~\bibnamefont {Wolf}},\ }\href {\doibase
  10.1103/PhysRevLett.99.197001} {\bibfield  {journal} {\bibinfo  {journal}
  {Phys. Rev. Lett.}\ }\textbf {\bibinfo {volume} {99}},\ \bibinfo {pages}
  {197001} (\bibinfo {year} {2007})}\BibitemShut {NoStop}%
\bibitem [{\citenamefont {Rameau}\ \emph {et~al.}(2014)\citenamefont {Rameau},
  \citenamefont {Freutel}, \citenamefont {Rettig}, \citenamefont {Avigo},
  \citenamefont {Ligges}, \citenamefont {Yoshida}, \citenamefont {Eisaki},
  \citenamefont {Schneeloch}, \citenamefont {Zhong}, \citenamefont {Xu},
  \citenamefont {Gu}, \citenamefont {Johnson},\ and\ \citenamefont
  {Bovensiepen}}]{rameautrarpes}%
  \BibitemOpen
  \bibfield  {author} {\bibinfo {author} {\bibfnamefont {J.~D.}\ \bibnamefont
  {Rameau}}, \bibinfo {author} {\bibfnamefont {S.}~\bibnamefont {Freutel}},
  \bibinfo {author} {\bibfnamefont {L.}~\bibnamefont {Rettig}}, \bibinfo
  {author} {\bibfnamefont {I.}~\bibnamefont {Avigo}}, \bibinfo {author}
  {\bibfnamefont {M.}~\bibnamefont {Ligges}}, \bibinfo {author} {\bibfnamefont
  {Y.}~\bibnamefont {Yoshida}}, \bibinfo {author} {\bibfnamefont
  {H.}~\bibnamefont {Eisaki}}, \bibinfo {author} {\bibfnamefont
  {J.}~\bibnamefont {Schneeloch}}, \bibinfo {author} {\bibfnamefont {R.~D.}\
  \bibnamefont {Zhong}}, \bibinfo {author} {\bibfnamefont {Z.~J.}\ \bibnamefont
  {Xu}}, \bibinfo {author} {\bibfnamefont {G.~D.}\ \bibnamefont {Gu}}, \bibinfo
  {author} {\bibfnamefont {P.~D.}\ \bibnamefont {Johnson}}, \ and\ \bibinfo
  {author} {\bibfnamefont {U.}~\bibnamefont {Bovensiepen}},\ }\href {\doibase
  10.1103/PhysRevB.89.115115} {\bibfield  {journal} {\bibinfo  {journal} {Phys.
  Rev. B}\ }\textbf {\bibinfo {volume} {89}},\ \bibinfo {pages} {115115}
  (\bibinfo {year} {2014})}\BibitemShut {NoStop}%
\bibitem [{\citenamefont {Guo}\ \emph {et~al.}(2011{\natexlab{a}})\citenamefont
  {Guo}, \citenamefont {Wulin}, \citenamefont {Chien},\ and\ \citenamefont
  {Levin}}]{levinprl}%
  \BibitemOpen
  \bibfield  {author} {\bibinfo {author} {\bibfnamefont {H.}~\bibnamefont
  {Guo}}, \bibinfo {author} {\bibfnamefont {D.}~\bibnamefont {Wulin}}, \bibinfo
  {author} {\bibfnamefont {C.-C.}\ \bibnamefont {Chien}}, \ and\ \bibinfo
  {author} {\bibfnamefont {K.}~\bibnamefont {Levin}},\ }\href {\doibase
  10.1103/PhysRevLett.107.020403} {\bibfield  {journal} {\bibinfo  {journal}
  {Phys. Rev. Lett.}\ }\textbf {\bibinfo {volume} {107}},\ \bibinfo {pages}
  {020403} (\bibinfo {year} {2011}{\natexlab{a}})}\BibitemShut {NoStop}%
\bibitem [{\citenamefont {Hartnoll}\ and\ \citenamefont
  {Hofman}(2012)}]{hartnoll3}%
  \BibitemOpen
  \bibfield  {author} {\bibinfo {author} {\bibfnamefont {S.~A.}\ \bibnamefont
  {Hartnoll}}\ and\ \bibinfo {author} {\bibfnamefont {D.~M.}\ \bibnamefont
  {Hofman}},\ }\href {\doibase 10.1103/PhysRevLett.108.241601} {\bibfield
  {journal} {\bibinfo  {journal} {Phys. Rev. Lett.}\ }\textbf {\bibinfo
  {volume} {108}},\ \bibinfo {pages} {241601} (\bibinfo {year}
  {2012})}\BibitemShut {NoStop}%
\bibitem [{\citenamefont {Allen}(1996)}]{allen}%
  \BibitemOpen
  \bibfield  {author} {\bibinfo {author} {\bibfnamefont {P.}~\bibnamefont
  {Allen}},\ }\enquote {\bibinfo {title} {Quantum theory of real materials},}\
  \ (\bibinfo  {publisher} {Kluwer},\ \bibinfo {address} {Boston},\ \bibinfo
  {year} {1996})\ pp.\ \bibinfo {pages} {219--250}\BibitemShut {NoStop}%
\bibitem [{\citenamefont {Varma}\ \emph {et~al.}(2002)\citenamefont {Varma},
  \citenamefont {Nussinov},\ and\ \citenamefont {van Saarloos}}]{varmalong}%
  \BibitemOpen
  \bibfield  {author} {\bibinfo {author} {\bibfnamefont {C.}~\bibnamefont
  {Varma}}, \bibinfo {author} {\bibfnamefont {Z.}~\bibnamefont {Nussinov}}, \
  and\ \bibinfo {author} {\bibfnamefont {W.}~\bibnamefont {van Saarloos}},\
  }\href {\doibase http://dx.doi.org/10.1016/S0370-1573(01)00060-6} {\bibfield
  {journal} {\bibinfo  {journal} {Physics Reports}\ }\textbf {\bibinfo {volume}
  {361}},\ \bibinfo {pages} {267 } (\bibinfo {year} {2002})}\BibitemShut
  {NoStop}%
\bibitem [{\citenamefont {Valla}\ \emph {et~al.}(1999)\citenamefont {Valla},
  \citenamefont {Fedorov}, \citenamefont {Johnson}, \citenamefont {Wells},
  \citenamefont {Hulbert}, \citenamefont {Li}, \citenamefont {Gu},\ and\
  \citenamefont {Koshizuka}}]{valla2}%
  \BibitemOpen
  \bibfield  {author} {\bibinfo {author} {\bibfnamefont {T.}~\bibnamefont
  {Valla}}, \bibinfo {author} {\bibfnamefont {A.~V.}\ \bibnamefont {Fedorov}},
  \bibinfo {author} {\bibfnamefont {P.~D.}\ \bibnamefont {Johnson}}, \bibinfo
  {author} {\bibfnamefont {B.~O.}\ \bibnamefont {Wells}}, \bibinfo {author}
  {\bibfnamefont {S.~L.}\ \bibnamefont {Hulbert}}, \bibinfo {author}
  {\bibfnamefont {Q.}~\bibnamefont {Li}}, \bibinfo {author} {\bibfnamefont
  {G.~D.}\ \bibnamefont {Gu}}, \ and\ \bibinfo {author} {\bibfnamefont
  {N.}~\bibnamefont {Koshizuka}},\ }\href {\doibase
  10.1126/science.285.5436.2110} {\bibfield  {journal} {\bibinfo  {journal}
  {Science}\ }\textbf {\bibinfo {volume} {285}},\ \bibinfo {pages} {2110}
  (\bibinfo {year} {1999})}\BibitemShut {NoStop}%
\bibitem [{\citenamefont {Vishik}\ \emph {et~al.}(2010)\citenamefont {Vishik},
  \citenamefont {Lee}, \citenamefont {Schmitt}, \citenamefont {Moritz},
  \citenamefont {Sasagawa}, \citenamefont {Uchida}, \citenamefont {Fujita},
  \citenamefont {Ishida}, \citenamefont {Zhang}, \citenamefont {Devereaux},\
  and\ \citenamefont {Shen}}]{vishik}%
  \BibitemOpen
  \bibfield  {author} {\bibinfo {author} {\bibfnamefont {I.~M.}\ \bibnamefont
  {Vishik}}, \bibinfo {author} {\bibfnamefont {W.~S.}\ \bibnamefont {Lee}},
  \bibinfo {author} {\bibfnamefont {F.}~\bibnamefont {Schmitt}}, \bibinfo
  {author} {\bibfnamefont {B.}~\bibnamefont {Moritz}}, \bibinfo {author}
  {\bibfnamefont {T.}~\bibnamefont {Sasagawa}}, \bibinfo {author}
  {\bibfnamefont {S.}~\bibnamefont {Uchida}}, \bibinfo {author} {\bibfnamefont
  {K.}~\bibnamefont {Fujita}}, \bibinfo {author} {\bibfnamefont
  {S.}~\bibnamefont {Ishida}}, \bibinfo {author} {\bibfnamefont
  {C.}~\bibnamefont {Zhang}}, \bibinfo {author} {\bibfnamefont {T.~P.}\
  \bibnamefont {Devereaux}}, \ and\ \bibinfo {author} {\bibfnamefont {Z.~X.}\
  \bibnamefont {Shen}},\ }\href {\doibase 10.1103/PhysRevLett.104.207002}
  {\bibfield  {journal} {\bibinfo  {journal} {Phys. Rev. Lett.}\ }\textbf
  {\bibinfo {volume} {104}},\ \bibinfo {pages} {207002} (\bibinfo {year}
  {2010})}\BibitemShut {NoStop}%
\bibitem [{\citenamefont {Doniach}\ and\ \citenamefont
  {Sondheimer}(1998)}]{greenfunbook}%
  \BibitemOpen
  \bibfield  {author} {\bibinfo {author} {\bibfnamefont {S.}~\bibnamefont
  {Doniach}}\ and\ \bibinfo {author} {\bibfnamefont {E.~H.}\ \bibnamefont
  {Sondheimer}},\ }\href@noop {} {\emph {\bibinfo {title} {Green's Functions
  for Solid State Physicists}}}\ (\bibinfo  {publisher} {Imperial College
  Press},\ \bibinfo {address} {London},\ \bibinfo {year} {1998})\BibitemShut
  {NoStop}%
\bibitem [{\citenamefont {Adams}\ \emph {et~al.}(2012)\citenamefont {Adams},
  \citenamefont {Carr}, \citenamefont {Sch\"{a}fer}, \citenamefont
  {Steinberg},\ and\ \citenamefont {Thomas}}]{njphys}%
  \BibitemOpen
  \bibfield  {author} {\bibinfo {author} {\bibfnamefont {A.}~\bibnamefont
  {Adams}}, \bibinfo {author} {\bibfnamefont {L.~D.}\ \bibnamefont {Carr}},
  \bibinfo {author} {\bibfnamefont {T.}~\bibnamefont {Sch\"{a}fer}}, \bibinfo
  {author} {\bibfnamefont {P.}~\bibnamefont {Steinberg}}, \ and\ \bibinfo
  {author} {\bibfnamefont {J.~E.}\ \bibnamefont {Thomas}},\ }\href {\doibase
  1367-2630/12/115009} {\bibfield  {journal} {\bibinfo  {journal} {New Journal
  of Physics}\ }\textbf {\bibinfo {volume} {14}},\ \bibinfo {pages} {115009}
  (\bibinfo {year} {2012})}\BibitemShut {NoStop}%
\bibitem [{\citenamefont {Huang}(1987)}]{huang}%
  \BibitemOpen
  \bibfield  {author} {\bibinfo {author} {\bibfnamefont {K.}~\bibnamefont
  {Huang}},\ }\href@noop {} {\emph {\bibinfo {title} {Statistical
  Mechanics}}},\ \bibinfo {edition} {2nd}\ ed.\ (\bibinfo  {publisher} {John
  Wiley and Sons},\ \bibinfo {address} {New York},\ \bibinfo {year}
  {1987})\BibitemShut {NoStop}%
\bibitem [{\citenamefont {Steinberg}(1958)}]{oldvisc}%
  \BibitemOpen
  \bibfield  {author} {\bibinfo {author} {\bibfnamefont {M.~S.}\ \bibnamefont
  {Steinberg}},\ }\href {\doibase 10.1103/PhysRev.109.1486} {\bibfield
  {journal} {\bibinfo  {journal} {Phys. Rev.}\ }\textbf {\bibinfo {volume}
  {109}},\ \bibinfo {pages} {1486} (\bibinfo {year} {1958})}\BibitemShut
  {NoStop}%
\bibitem [{\citenamefont {Guo}\ \emph {et~al.}(2011{\natexlab{b}})\citenamefont
  {Guo}, \citenamefont {Wulin}, \citenamefont {Chien},\ and\ \citenamefont
  {Levin}}]{klevin}%
  \BibitemOpen
  \bibfield  {author} {\bibinfo {author} {\bibfnamefont {H.}~\bibnamefont
  {Guo}}, \bibinfo {author} {\bibfnamefont {D.}~\bibnamefont {Wulin}}, \bibinfo
  {author} {\bibfnamefont {C.-C.}\ \bibnamefont {Chien}}, \ and\ \bibinfo
  {author} {\bibfnamefont {K.}~\bibnamefont {Levin}},\ }\href {\doibase
  10.1088/1367-2630/13/7/075011} {\bibfield  {journal} {\bibinfo  {journal}
  {New Journal of Physics}\ }\textbf {\bibinfo {volume} {13}},\ \bibinfo
  {pages} {075011} (\bibinfo {year} {2011}{\natexlab{b}})}\BibitemShut
  {NoStop}%
\bibitem [{\citenamefont {M\"uller}\ \emph {et~al.}(2009)\citenamefont
  {M\"uller}, \citenamefont {Schmalian},\ and\ \citenamefont
  {Fritz}}]{schmalian}%
  \BibitemOpen
  \bibfield  {author} {\bibinfo {author} {\bibfnamefont {M.}~\bibnamefont
  {M\"uller}}, \bibinfo {author} {\bibfnamefont {J.}~\bibnamefont {Schmalian}},
  \ and\ \bibinfo {author} {\bibfnamefont {L.}~\bibnamefont {Fritz}},\ }\href
  {\doibase 10.1103/PhysRevLett.103.025301} {\bibfield  {journal} {\bibinfo
  {journal} {Phys. Rev. Lett.}\ }\textbf {\bibinfo {volume} {103}},\ \bibinfo
  {pages} {025301} (\bibinfo {year} {2009})}\BibitemShut {NoStop}%
\bibitem [{\citenamefont {Ashcroft}\ and\ \citenamefont
  {Mermin}(1976)}]{ashcroft}%
  \BibitemOpen
  \bibfield  {author} {\bibinfo {author} {\bibfnamefont {N.}~\bibnamefont
  {Ashcroft}}\ and\ \bibinfo {author} {\bibfnamefont {N.}~\bibnamefont
  {Mermin}},\ }\href@noop {} {\emph {\bibinfo {title} {Solid State Physics}}}\
  (\bibinfo  {publisher} {Saunders College},\ \bibinfo {address}
  {Philadelphia},\ \bibinfo {year} {1976})\BibitemShut {NoStop}%
\bibitem [{\citenamefont {Yang}\ \emph {et~al.}(2006)\citenamefont {Yang},
  \citenamefont {Rice},\ and\ \citenamefont {Zhang}}]{yrz}%
  \BibitemOpen
  \bibfield  {author} {\bibinfo {author} {\bibfnamefont {K.-Y.}\ \bibnamefont
  {Yang}}, \bibinfo {author} {\bibfnamefont {T.~M.}\ \bibnamefont {Rice}}, \
  and\ \bibinfo {author} {\bibfnamefont {F.-C.}\ \bibnamefont {Zhang}},\
  }\href@noop {} {\bibfield  {journal} {\bibinfo  {journal} {Phys. Rev. B}\
  }\textbf {\bibinfo {volume} {73}},\ \bibinfo {pages} {174501} (\bibinfo
  {year} {2006})}\BibitemShut {NoStop}%
\bibitem [{\citenamefont {Rameau}\ \emph {et~al.}(2010)\citenamefont {Rameau},
  \citenamefont {Yang},\ and\ \citenamefont {Johnson}}]{jonelecspec}%
  \BibitemOpen
  \bibfield  {author} {\bibinfo {author} {\bibfnamefont {J.}~\bibnamefont
  {Rameau}}, \bibinfo {author} {\bibfnamefont {H.-B.}\ \bibnamefont {Yang}}, \
  and\ \bibinfo {author} {\bibfnamefont {P.}~\bibnamefont {Johnson}},\ }\href
  {\doibase http://dx.doi.org/10.1016/j.elspec.2010.05.025} {\bibfield
  {journal} {\bibinfo  {journal} {Journal of Electron Spectroscopy and Related
  Phenomena}\ }\textbf {\bibinfo {volume} {181}},\ \bibinfo {pages} {35 }
  (\bibinfo {year} {2010})}\BibitemShut {NoStop}%
\bibitem [{\citenamefont {Norman}\ \emph {et~al.}(2000)\citenamefont {Norman},
  \citenamefont {Randeria}, \citenamefont {Jank\'{o}},\ and\ \citenamefont
  {Campuzano}}]{randeria2}%
  \BibitemOpen
  \bibfield  {author} {\bibinfo {author} {\bibfnamefont {M.~R.}\ \bibnamefont
  {Norman}}, \bibinfo {author} {\bibfnamefont {M.}~\bibnamefont {Randeria}},
  \bibinfo {author} {\bibfnamefont {B.}~\bibnamefont {Jank\'{o}}}, \ and\
  \bibinfo {author} {\bibfnamefont {J.~C.}\ \bibnamefont {Campuzano}},\
  }\href@noop {} {\bibfield  {journal} {\bibinfo  {journal} {Phys. Rev. B}\
  }\textbf {\bibinfo {volume} {61}},\ \bibinfo {pages} {14742} (\bibinfo {year}
  {2000})}\BibitemShut {NoStop}%
\bibitem [{\citenamefont {Yang}\ \emph {et~al.}(2008)\citenamefont {Yang},
  \citenamefont {Rameau}, \citenamefont {Johnson}, \citenamefont {Valla},
  \citenamefont {Tsvelik},\ and\ \citenamefont {Gu}}]{yang2}%
  \BibitemOpen
  \bibfield  {author} {\bibinfo {author} {\bibfnamefont {H.-B.}\ \bibnamefont
  {Yang}}, \bibinfo {author} {\bibfnamefont {J.~D.}\ \bibnamefont {Rameau}},
  \bibinfo {author} {\bibfnamefont {P.~D.}\ \bibnamefont {Johnson}}, \bibinfo
  {author} {\bibfnamefont {T.}~\bibnamefont {Valla}}, \bibinfo {author}
  {\bibfnamefont {A.}~\bibnamefont {Tsvelik}}, \ and\ \bibinfo {author}
  {\bibfnamefont {G.~D.}\ \bibnamefont {Gu}},\ }\href@noop {} {\bibfield
  {journal} {\bibinfo  {journal} {Nature (London)}\ }\textbf {\bibinfo {volume}
  {456}},\ \bibinfo {pages} {77} (\bibinfo {year} {2008})}\BibitemShut
  {NoStop}%
\bibitem [{\citenamefont {Hartnoll}\ \emph {et~al.}(2007)\citenamefont
  {Hartnoll}, \citenamefont {Kovtun}, \citenamefont {M\"uller},\ and\
  \citenamefont {Sachdev}}]{hartnollimpurity}%
  \BibitemOpen
  \bibfield  {author} {\bibinfo {author} {\bibfnamefont {S.~A.}\ \bibnamefont
  {Hartnoll}}, \bibinfo {author} {\bibfnamefont {P.~K.}\ \bibnamefont
  {Kovtun}}, \bibinfo {author} {\bibfnamefont {M.}~\bibnamefont {M\"uller}}, \
  and\ \bibinfo {author} {\bibfnamefont {S.}~\bibnamefont {Sachdev}},\ }\href
  {\doibase 10.1103/PhysRevB.76.144502} {\bibfield  {journal} {\bibinfo
  {journal} {Phys. Rev. B}\ }\textbf {\bibinfo {volume} {76}},\ \bibinfo
  {pages} {144502} (\bibinfo {year} {2007})}\BibitemShut {NoStop}%
\bibitem [{\citenamefont {Das~Sarma}\ and\ \citenamefont
  {Stern}(1985)}]{dassarma}%
  \BibitemOpen
  \bibfield  {author} {\bibinfo {author} {\bibfnamefont {S.}~\bibnamefont
  {Das~Sarma}}\ and\ \bibinfo {author} {\bibfnamefont {F.}~\bibnamefont
  {Stern}},\ }\href {\doibase 10.1103/PhysRevB.32.8442} {\bibfield  {journal}
  {\bibinfo  {journal} {Physical Review B}\ }\textbf {\bibinfo {volume} {32}},\
  \bibinfo {pages} {8442} (\bibinfo {year} {1985})}\BibitemShut {NoStop}%
\bibitem [{\citenamefont {Valla}\ \emph {et~al.}(2000)\citenamefont {Valla},
  \citenamefont {Fedorov}, \citenamefont {Johnson}, \citenamefont {Xue},
  \citenamefont {Smith},\ and\ \citenamefont {DiSalvo}}]{tasetwo}%
  \BibitemOpen
  \bibfield  {author} {\bibinfo {author} {\bibfnamefont {T.}~\bibnamefont
  {Valla}}, \bibinfo {author} {\bibfnamefont {A.~V.}\ \bibnamefont {Fedorov}},
  \bibinfo {author} {\bibfnamefont {P.~D.}\ \bibnamefont {Johnson}}, \bibinfo
  {author} {\bibfnamefont {J.}~\bibnamefont {Xue}}, \bibinfo {author}
  {\bibfnamefont {K.~E.}\ \bibnamefont {Smith}}, \ and\ \bibinfo {author}
  {\bibfnamefont {F.~J.}\ \bibnamefont {DiSalvo}},\ }\href {\doibase
  10.1103/PhysRevLett.85.4759} {\bibfield  {journal} {\bibinfo  {journal}
  {Phys. Rev. Lett.}\ }\textbf {\bibinfo {volume} {85}},\ \bibinfo {pages}
  {4759} (\bibinfo {year} {2000})}\BibitemShut {NoStop}%
\bibitem [{\citenamefont {Perfetti}\ \emph {et~al.}(2001)\citenamefont
  {Perfetti}, \citenamefont {Rojas}, \citenamefont {Reginelli}, \citenamefont
  {Gavioli}, \citenamefont {Berger}, \citenamefont {Margaritondo},
  \citenamefont {Grioni}, \citenamefont {Ga\'al}, \citenamefont {Forr\'o},\
  and\ \citenamefont {Rullier~Albenque}}]{perfettititetwo}%
  \BibitemOpen
  \bibfield  {author} {\bibinfo {author} {\bibfnamefont {L.}~\bibnamefont
  {Perfetti}}, \bibinfo {author} {\bibfnamefont {C.}~\bibnamefont {Rojas}},
  \bibinfo {author} {\bibfnamefont {A.}~\bibnamefont {Reginelli}}, \bibinfo
  {author} {\bibfnamefont {L.}~\bibnamefont {Gavioli}}, \bibinfo {author}
  {\bibfnamefont {H.}~\bibnamefont {Berger}}, \bibinfo {author} {\bibfnamefont
  {G.}~\bibnamefont {Margaritondo}}, \bibinfo {author} {\bibfnamefont
  {M.}~\bibnamefont {Grioni}}, \bibinfo {author} {\bibfnamefont
  {R.}~\bibnamefont {Ga\'al}}, \bibinfo {author} {\bibfnamefont
  {L.}~\bibnamefont {Forr\'o}}, \ and\ \bibinfo {author} {\bibfnamefont
  {F.}~\bibnamefont {Rullier~Albenque}},\ }\href {\doibase
  10.1103/PhysRevB.64.115102} {\bibfield  {journal} {\bibinfo  {journal} {Phys.
  Rev. B}\ }\textbf {\bibinfo {volume} {64}},\ \bibinfo {pages} {115102}
  (\bibinfo {year} {2001})}\BibitemShut {NoStop}%
\bibitem [{\citenamefont {Valla}\ \emph {et~al.}(2006)\citenamefont {Valla},
  \citenamefont {Kidd}, \citenamefont {Rameau}, \citenamefont {Noh},
  \citenamefont {Gu}, \citenamefont {Johnson}, \citenamefont {Yang},\ and\
  \citenamefont {Ding}}]{finedetails}%
  \BibitemOpen
  \bibfield  {author} {\bibinfo {author} {\bibfnamefont {T.}~\bibnamefont
  {Valla}}, \bibinfo {author} {\bibfnamefont {T.~E.}\ \bibnamefont {Kidd}},
  \bibinfo {author} {\bibfnamefont {J.~D.}\ \bibnamefont {Rameau}}, \bibinfo
  {author} {\bibfnamefont {H.-J.}\ \bibnamefont {Noh}}, \bibinfo {author}
  {\bibfnamefont {G.~D.}\ \bibnamefont {Gu}}, \bibinfo {author} {\bibfnamefont
  {P.~D.}\ \bibnamefont {Johnson}}, \bibinfo {author} {\bibfnamefont {H.-B.}\
  \bibnamefont {Yang}}, \ and\ \bibinfo {author} {\bibfnamefont
  {H.}~\bibnamefont {Ding}},\ }\href {\doibase 10.1103/PhysRevB.73.184518}
  {\bibfield  {journal} {\bibinfo  {journal} {Phys. Rev. B}\ }\textbf {\bibinfo
  {volume} {73}},\ \bibinfo {pages} {184518} (\bibinfo {year}
  {2006})}\BibitemShut {NoStop}%
\bibitem [{\citenamefont {Hwang}\ \emph {et~al.}(2004)\citenamefont {Hwang},
  \citenamefont {Timusk},\ and\ \citenamefont {Gu}}]{opticond}%
  \BibitemOpen
  \bibfield  {author} {\bibinfo {author} {\bibfnamefont {J.}~\bibnamefont
  {Hwang}}, \bibinfo {author} {\bibfnamefont {T.}~\bibnamefont {Timusk}}, \
  and\ \bibinfo {author} {\bibfnamefont {G.}~\bibnamefont {Gu}},\ }\href
  {\doibase doi:10.1038/nature02347} {\bibfield  {journal} {\bibinfo  {journal}
  {Nature (London)}\ }\textbf {\bibinfo {volume} {427}},\ \bibinfo {pages}
  {714} (\bibinfo {year} {2004})}\BibitemShut {NoStop}%
\bibitem [{\citenamefont {Plumb}\ \emph {et~al.}(2010)\citenamefont {Plumb},
  \citenamefont {Reber}, \citenamefont {Koralek}, \citenamefont {Sun},
  \citenamefont {Douglas}, \citenamefont {Aiura}, \citenamefont {Oka},
  \citenamefont {Eisaki},\ and\ \citenamefont {Dessau}}]{plumb}%
  \BibitemOpen
  \bibfield  {author} {\bibinfo {author} {\bibfnamefont {N.~C.}\ \bibnamefont
  {Plumb}}, \bibinfo {author} {\bibfnamefont {T.~J.}\ \bibnamefont {Reber}},
  \bibinfo {author} {\bibfnamefont {J.~D.}\ \bibnamefont {Koralek}}, \bibinfo
  {author} {\bibfnamefont {Z.}~\bibnamefont {Sun}}, \bibinfo {author}
  {\bibfnamefont {J.~F.}\ \bibnamefont {Douglas}}, \bibinfo {author}
  {\bibfnamefont {Y.}~\bibnamefont {Aiura}}, \bibinfo {author} {\bibfnamefont
  {K.}~\bibnamefont {Oka}}, \bibinfo {author} {\bibfnamefont {H.}~\bibnamefont
  {Eisaki}}, \ and\ \bibinfo {author} {\bibfnamefont {D.~S.}\ \bibnamefont
  {Dessau}},\ }\href {\doibase 10.1103/PhysRevLett.105.046402} {\bibfield
  {journal} {\bibinfo  {journal} {Phys. Rev. Lett.}\ }\textbf {\bibinfo
  {volume} {105}},\ \bibinfo {pages} {046402} (\bibinfo {year}
  {2010})}\BibitemShut {NoStop}%
\bibitem [{\citenamefont {Pines}\ and\ \citenamefont
  {Nozi\`{e}res}(1966)}]{pines}%
  \BibitemOpen
  \bibfield  {author} {\bibinfo {author} {\bibfnamefont {D.}~\bibnamefont
  {Pines}}\ and\ \bibinfo {author} {\bibfnamefont {P.}~\bibnamefont
  {Nozi\`{e}res}},\ }\href@noop {} {\emph {\bibinfo {title} {The Theory of
  Quantum Liquids}}}\ (\bibinfo  {publisher} {W.A. Benjamin, Inc.},\ \bibinfo
  {address} {New York, New York, 10019},\ \bibinfo {year} {1966})\BibitemShut
  {NoStop}%
\bibitem [{\citenamefont {\mbox{PHENIX} Collaboration}(2012)}]{sphenix}%
  \BibitemOpen
  \bibfield  {author} {\bibinfo {author} {\bibnamefont {\mbox{PHENIX}
  Collaboration}},\ }\href@noop {} {\bibfield  {journal} {\bibinfo  {journal}
  {arXiv:1207.6378v2}\ } (\bibinfo {year} {2012})}\BibitemShut {NoStop}%
\bibitem [{\citenamefont {Cau}\ \emph {et~al.}(2011)\citenamefont {Cau},
  \citenamefont {Elliott}, \citenamefont {Joseph}, \citenamefont {Wu},
  \citenamefont {Petricka}, \citenamefont {Sch\"{a}fer},\ and\ \citenamefont
  {Thomas}}]{UFG1}%
  \BibitemOpen
  \bibfield  {author} {\bibinfo {author} {\bibfnamefont {C.}~\bibnamefont
  {Cau}}, \bibinfo {author} {\bibfnamefont {E.}~\bibnamefont {Elliott}},
  \bibinfo {author} {\bibfnamefont {J.}~\bibnamefont {Joseph}}, \bibinfo
  {author} {\bibfnamefont {J.}~\bibnamefont {Wu}}, \bibinfo {author}
  {\bibfnamefont {J.}~\bibnamefont {Petricka}}, \bibinfo {author}
  {\bibfnamefont {T.}~\bibnamefont {Sch\"{a}fer}}, \ and\ \bibinfo {author}
  {\bibfnamefont {J.~E.}\ \bibnamefont {Thomas}},\ }\href {\doibase
  10.1126/science.1195219} {\bibfield  {journal} {\bibinfo  {journal}
  {Science}\ }\textbf {\bibinfo {volume} {331}},\ \bibinfo {pages} {58}
  (\bibinfo {year} {2011})}\BibitemShut {NoStop}%
\bibitem [{\citenamefont {Tallon}\ \emph {et~al.}(2003)\citenamefont {Tallon},
  \citenamefont {Loram}, \citenamefont {Cooper}, \citenamefont {Panagopoulos},\
  and\ \citenamefont {Bernhard}}]{Tallon1}%
  \BibitemOpen
  \bibfield  {author} {\bibinfo {author} {\bibfnamefont {J.~L.}\ \bibnamefont
  {Tallon}}, \bibinfo {author} {\bibfnamefont {J.~W.}\ \bibnamefont {Loram}},
  \bibinfo {author} {\bibfnamefont {J.~R.}\ \bibnamefont {Cooper}}, \bibinfo
  {author} {\bibfnamefont {C.}~\bibnamefont {Panagopoulos}}, \ and\ \bibinfo
  {author} {\bibfnamefont {C.}~\bibnamefont {Bernhard}},\ }\href {\doibase
  10.1103/PhysRevB.68.180501} {\bibfield  {journal} {\bibinfo  {journal} {Phys.
  Rev. B}\ }\textbf {\bibinfo {volume} {68}},\ \bibinfo {pages} {180501}
  (\bibinfo {year} {2003})}\BibitemShut {NoStop}%
\bibitem [{\citenamefont {Gale}\ \emph {et~al.}(2013)\citenamefont {Gale},
  \citenamefont {Jeon}, \citenamefont {Schenke}, \citenamefont {Tribedy},\ and\
  \citenamefont {Venugopalan}}]{glasma}%
  \BibitemOpen
  \bibfield  {author} {\bibinfo {author} {\bibfnamefont {C.}~\bibnamefont
  {Gale}}, \bibinfo {author} {\bibfnamefont {S.}~\bibnamefont {Jeon}}, \bibinfo
  {author} {\bibfnamefont {B.}~\bibnamefont {Schenke}}, \bibinfo {author}
  {\bibfnamefont {P.}~\bibnamefont {Tribedy}}, \ and\ \bibinfo {author}
  {\bibfnamefont {R.}~\bibnamefont {Venugopalan}},\ }\href {\doibase
  10.1103/PhysRevLett.110.012302} {\bibfield  {journal} {\bibinfo  {journal}
  {Phys. Rev. Lett.}\ }\textbf {\bibinfo {volume} {110}},\ \bibinfo {pages}
  {012302} (\bibinfo {year} {2013})}\BibitemShut {NoStop}%
\bibitem [{\citenamefont {Zaanen}(2004)}]{zaanenplanck}%
  \BibitemOpen
  \bibfield  {author} {\bibinfo {author} {\bibfnamefont {J.}~\bibnamefont
  {Zaanen}},\ }\href@noop {} {\bibfield  {journal} {\bibinfo  {journal} {Nature
  (London)}\ }\textbf {\bibinfo {volume} {430}},\ \bibinfo {pages} {512}
  (\bibinfo {year} {2004})}\BibitemShut {NoStop}%
\bibitem [{\citenamefont {Song}\ \emph {et~al.}(2011)\citenamefont {Song},
  \citenamefont {Bass}, \citenamefont {Heinz}, \citenamefont {Hirano},\ and\
  \citenamefont {Shen}}]{prcqgp}%
  \BibitemOpen
  \bibfield  {author} {\bibinfo {author} {\bibfnamefont {H.}~\bibnamefont
  {Song}}, \bibinfo {author} {\bibfnamefont {S.~A.}\ \bibnamefont {Bass}},
  \bibinfo {author} {\bibfnamefont {U.}~\bibnamefont {Heinz}}, \bibinfo
  {author} {\bibfnamefont {T.}~\bibnamefont {Hirano}}, \ and\ \bibinfo {author}
  {\bibfnamefont {C.}~\bibnamefont {Shen}},\ }\href {\doibase
  10.1103/PhysRevC.83.054910} {\bibfield  {journal} {\bibinfo  {journal} {Phys.
  Rev. C}\ }\textbf {\bibinfo {volume} {83}},\ \bibinfo {pages} {054910}
  (\bibinfo {year} {2011})}\BibitemShut {NoStop}%
\bibitem [{\citenamefont {Schwabl}(2002)}]{schwabl}%
  \BibitemOpen
  \bibfield  {author} {\bibinfo {author} {\bibfnamefont {F.}~\bibnamefont
  {Schwabl}},\ }\href@noop {} {\emph {\bibinfo {title} {Statistical
  Mechanics}}}\ (\bibinfo  {publisher} {Springer-Verlag},\ \bibinfo {address}
  {Berlin, Germany},\ \bibinfo {year} {2002})\BibitemShut {NoStop}%
\bibitem [{\citenamefont {Yang}\ \emph {et~al.}(2009)\citenamefont {Yang},
  \citenamefont {Yang}, \citenamefont {Johnson}, \citenamefont {Rice},\ and\
  \citenamefont {Zhang}}]{yrzandrewcalc}%
  \BibitemOpen
  \bibfield  {author} {\bibinfo {author} {\bibfnamefont {K.-Y.}\ \bibnamefont
  {Yang}}, \bibinfo {author} {\bibfnamefont {H.-B.}\ \bibnamefont {Yang}},
  \bibinfo {author} {\bibfnamefont {P.}~\bibnamefont {Johnson}}, \bibinfo
  {author} {\bibfnamefont {T.}~\bibnamefont {Rice}}, \ and\ \bibinfo {author}
  {\bibfnamefont {F.-C.}\ \bibnamefont {Zhang}},\ }\href@noop {} {\bibfield
  {journal} {\bibinfo  {journal} {Europhysics Letters}\ }\textbf {\bibinfo
  {volume} {86}},\ \bibinfo {pages} {37002} (\bibinfo {year}
  {2009})}\BibitemShut {NoStop}%
\end{thebibliography}
\end{document}